\documentclass[12pt,preprint]{aastex}

\usepackage{graphicx}
\usepackage{multicol}
\usepackage{amsmath}
\usepackage{float}
\providecommand{\e}[1]{\ensuremath{\times 10^{#1}}}
\newcommand{\Msun}{\ensuremath{M_{\odot}}}

\shorttitle{Stellar Populations in Outer Disks}
\shortauthors{Alberts et al.}

\begin{document}

\title{The Evolution of Stellar Populations in the Outer Disks of Spiral Galaxies\altaffilmark{1}}

\author{Stacey Alberts\altaffilmark{2}, Daniela Calzetti\altaffilmark{2}, Hui Dong\altaffilmark{2}, L. C. Johnson\altaffilmark{3}, Daniel A. Dale\altaffilmark{4}, Luciana Bianchi\altaffilmark{5}, Rupali Chandar\altaffilmark{6}, Robert C. Kennicutt\altaffilmark{7}, Gerhardt R. Meurer\altaffilmark{8}, Michael Regan\altaffilmark{9}, David Thilker\altaffilmark{5}}

\altaffiltext{1}{Based on observations obtained with the Spitzer Space Telescope as
part of the Cycle 3 GO program 30753.}
\altaffiltext{2}{Department of Astronomy, LGRT-B 619E, University of Massachusetts, Amherst, MA 01003, USA} 
\altaffiltext{3}{Department of Astronomy, University of Washington, Seattle, WA 98195, USA}
\altaffiltext{4}{Department of Physics and Astronomy, University of Wyoming, Laramie, WY 82071, USA} 
\altaffiltext{5}{Department of Physics and Astronomy, Johns Hopkins University, Baltimore, MD 21218, USA}
\altaffiltext{6}{Department of Physics and Astronomy, University of Toledo, Toledo, OH 43606, USA}
\altaffiltext{7}{Institute of Astronomy, Cambridge University, Cambridge CB3 0HA, UK}
\altaffiltext{8}{International Centre for Radio Astronomy, University of Western Australia, Perth, WA 6009, AUS}
\altaffiltext{9}{Space Telescope Science Institute, Baltimore, MD 21218, USA}

\begin{abstract}
We investigate recent star formation in the extended ultraviolet (XUV) disks of five nearby galaxies (NGC 0628, NGC 2090, NGC 2841, NGC 3621, and NGC 5055) using a long wavelength baseline comprised of ultraviolet and mid-infrared imaging from the Galaxy Evolution Explorer and the Spitzer Infrared Array Camera.  We identify 229 unresolved stellar complexes across targeted portions of their XUV disks and utilize spectral energy distribution fitting to measure their stellar ages and masses through comparison with Starburst99 population synthesis models of instantaneous burst populations.  We find that the median age of outer disk associations in our sample is $\sim$100 Myr with a large dispersion that spans the entire range of our models (1 Myr$-$1 Gyr).  This relatively evolved state for most associations addresses the observed dearth of H$\alpha$ emission in some outer disks, as H$\alpha$ can only be observed in star forming regions younger than $\sim$10 Myr.  The large age dispersion is robust against variations in extinction (in the range $E(B-V)$=0$-$0.3 mag) and variations in the upper end of the stellar Initial Mass Function (IMF).  In particular, we demonstrate that the age dispersion is insensitive to steepening of the IMF, up to extreme slopes.


\end{abstract}


\section{Introduction}

The discovery of in-situ star formation (SF) in the extended ultraviolet (XUV) disks of spiral galaxies \citep{thi07} offers us a chance to study the processes of star formation in extreme low-density environments.  As the driving force behind the formation and evolution of galactic disks, star formation has been extensively studied in the high-density `inner' disk.  Extending this understanding with studies in the outer disk will have many far-reaching implications.  Are we in the presence of a different mode of star formation relative to that of the inner disks?  Is the stellar Initial Mass Function (IMF) observed in inner disks applicable to outer disk environments?  Can we confirm or challenge current models or scaling relations such as the Schmidt-Kennicutt law \citep{ken98a} or the classical SF threshold \citep{ken89,mar01}?  What effects do the feedback processes from massive stars in the outer disks have on, and are they a primary component in, the chemical enrichment of the pristine halo?  Star formation in the XUV disks of nearby galaxies can be studied in great detail and, as SF is found to occur later in the outer disk than the inner disk \citep{mun07,rok08}, it may provide a similar environment to the low-density, low-metallicity, pristine conditions of newly forming galaxies in the early Universe.  In addition, the long term implications of galaxy interactions may still be imprinted on these XUV disks.

The possibility of star formation in the outer disks of galaxies was first revealed by the presence of H$\alpha$ emission out beyond the galaxies' optical radii \citep{fer98,lel00} and by the discovery of B stars in the outer disk of M31 \citep{cui01}.  This was confirmed by the Galaxy Evolution Explorer (GALEX), which discovered smooth ultraviolet (UV) profiles that extended well beyond $R_{25}$ in a number of galaxies \citep{thi05,gil05,boi07}.  This was quickly followed by a study of about 200 S0-Sm galaxies from the GALEX Nearby Galaxies Survey (NGS, e.g. \citet{bia03, bia09} and references therein; \citet{gil07b}), with corollary optical-NIR imaging by \citet{thi07}, who found XUV disks in $\sim$30$\%$ of local ($<$ 40 Mpc) galaxies and grouped them into two major types (see, also, \citet{zar07}).  These disks are characterized by the presence of many UV-bright knots \citep{thi05,gil05}, which have since been associated with HI structures, display metallicities in the range of Z/Z$_{\odot}$ = 1/10 $-$ 1 \citep{fer98, gil07a, bre09, wer10}, and have been determined to be locales of star forming activity.

As star formation tracers H$\alpha$, far-ultraviolet (FUV), and near-ultraviolet (NUV) emission all tell us about different populations \citep{ken98b}.  H$\alpha$ traces the ionizing population ($<$ 10 Myr), while the UV is composed of emission from all young stars, with FUV sensitive to the youngest phases of SF and NUV to a slightly more evolved population ($\sim$100 Myr).  It has been found that the UV emission is more extended than the H$\alpha$ emission in about half of XUV disks, with the remaining half showing roughly equal extent for UV and H$\alpha$ \citep{god10}. The cases of M83 and NGC 4625 demonstrate this dearth of H$\alpha$ emission, with only $\sim$10$-$20$\%$ of UV clumps showing associated H$\alpha$ \citep{thi05} and, in general, the H$\alpha$ spatial covering fraction recovered in these XUV disks is about half that seen in the UV \citep{god10}. We address two of the possible explanations that have been put forth: (1) the star forming associations in these regions exist at a wide range of ages with a  majority of associations beyond the age where ionizing stars still exist ($\sim$10 Myr) \citep{thi05,zar07,god10} or (2) in low-density regions high mass star formation is effectively suppressed via a top-light IMF \citep{meu09}.  These two solutions are generally difficult to separate due to the age-IMF degeneracy. In this paper, we aim to determine the age range and age distribution of star forming associations in the outer disks of five galaxies assuming a standard Kroupa IMF \citep{kro01} and utilizing a long wavelength baseline by combining GALEX far-ultraviolet and near-ultraviolet imaging with Spitzer mid-infrared (MIR; 3.6$-$8.0 $\mu$m) bands.  Then we will perform the same analysis using a top-light IMF and will outline the effects of steepening the IMF slope on the age/mass distribution of our associations.  The advantage of our approach, relative to previous studies, is that we do not rely on the presence of ionized gas to assess the existence of IMF variations; we instead exploit the long wavelength baseline afforded by the combination of GALEX and Spitzer data to break the age-IMF degeneracy.

In Section 2, we provide a brief description of how we selected our galaxies and the data used in this paper.  Section 3 outlines the process of association identification and selection as well as our photometry.  In Section 4, we discuss our models and how they are applied to the data.  In Section 5, we present the statistical results per galaxy of the age/mass distribution and, in Section 6, the analysis of how the IMF affects this distribution.  Section 7 concludes with a brief outline of future work.

\section{Galaxy Selection, Observations, and Data Reduction}

A recent study of M83 has demonstrated the effectiveness of combining the UV data from GALEX with the MIR bands of the Spitzer Infrared Array Camera (IRAC) to constrain the ages and masses of star forming associations \citep{don08}.  To expand on this study, five galaxies from the GALEX Atlas of Nearby Galaxies \citep{gil07b} were chosen as a sub-sample of XUV disks that represent the full range in UV morphologies.  These different morphologies have been hypothesised to be different phases of disk building \citep{thi07} and could indicate some  influence of the parent galaxy on outer disk populations. Target regions beyond 1.5 $R_{opt}$ were then chosen for Spitzer and imaged using IRAC's four bands: 3.6 $\mu$m, 4.5 $\mu$m, 5.8 $\mu$m and 8.0 $\mu$m.  These data join the FUV ($\sim$1529 $\AA$) and NUV ($\sim$2312 $\AA$) bands from GALEX.

\begin{figure}
\epsscale{0.7}\plotone{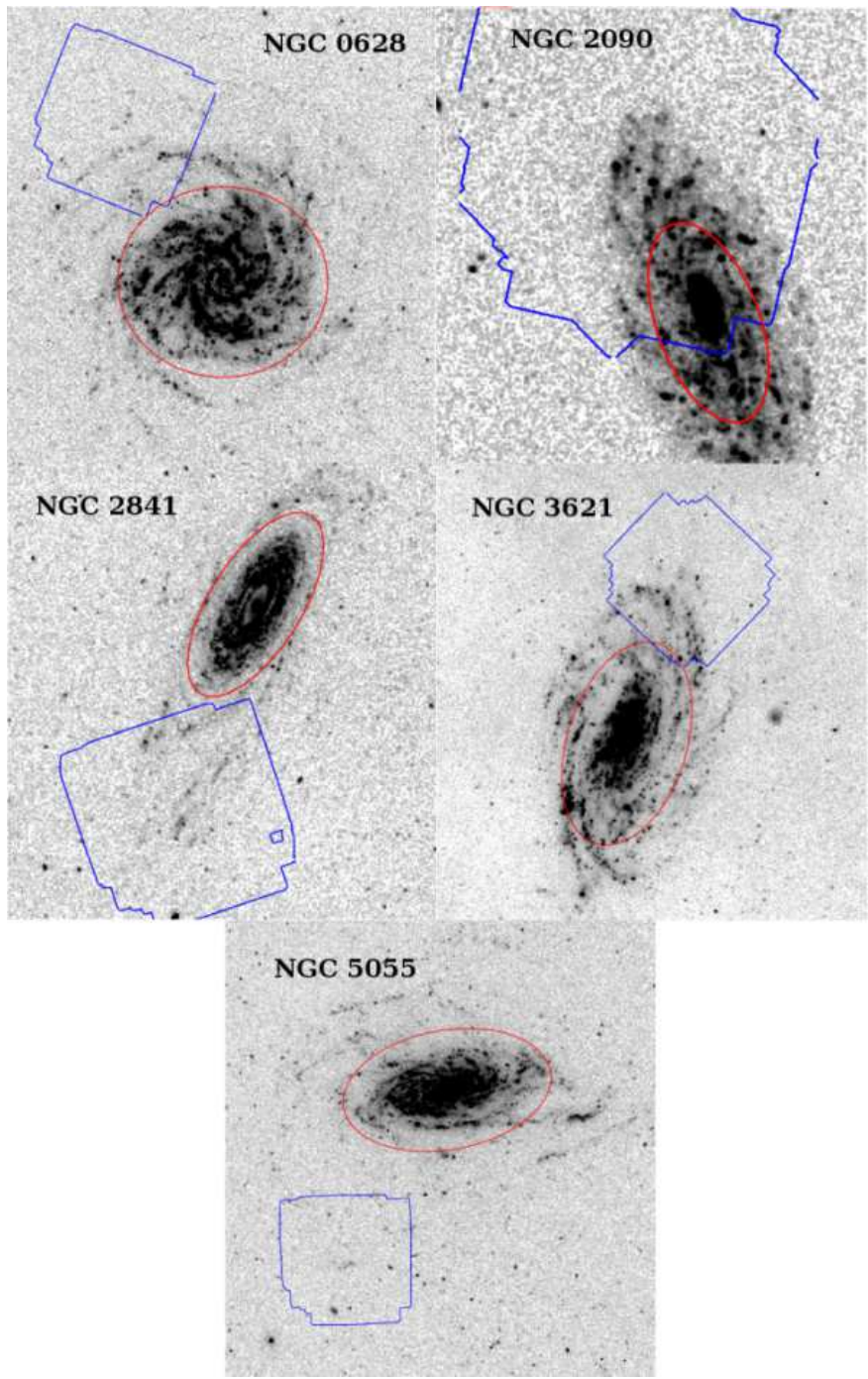}
\caption{The GALEX FUV images for the five galaxies in this study.  The red circle traces the R$_{25}$ boundary and the blue box shows the target region imaged with Spitzer/IRAC.}
\label{fig:whole}
\end{figure}

The five galaxies, NGC 0628, NGC 2090, NGC 2841, NGC 3621, and NGC 5055, can be seen in Figure~\ref{fig:whole}; Table~\ref{tbl:dist} contains their relevant information.  These galaxies were selected to be observed with Spitzer because they are close enough (6$-$12 Mpc) that the IRAC resolution of $\sim$2'', which corresponds to a physical scale of 60$-$110 pc, is larger than the typical size of a compact stellar cluster ($\sim$2$-$5 pc), but not much larger than the typical size of a scaled OB association, $\sim$50 pc \citep{mai01}.  The outer disk regions targeted with Spitzer were selected to have concentrated UV emission within the central $\sim$1~arcminute of the $\sim$5~arcminute Spitzer IRAC field of view in order to maximize the number of detections in the infrared. 



\begin{deluxetable}{ccccc}
  \tabletypesize{\scriptsize}
  \tablecaption{Galaxy Information}
  \tablewidth{0pt}
  \tablehead{
  \colhead{} &
  \colhead{RA} &
  \colhead{Dec} &
  \colhead{Distance} &
  \colhead{XUV-Disk} \\
  \noalign{\smallskip}
  \colhead{} &
  \colhead{(J2000)} &
  \colhead{(J2000)} &
  \colhead{(Mpc)} &
  \colhead{Type}
  }
  \startdata
  NGC 0628 & 01 36 41.77 & +15 47 00.50  & 7.3 & I \\
  NGC 2090 & 05 47 01.89 & $-$34 15 02.20  & 11.2 & II \\
  NGC 2841 & 09 22 02.63  & +50 58 35.47  & 9.8 & I \\
  NGC 3621 & 11 18 16.51 & $-$32 48 50.60  & 6.6 & I \\
  NGC 5055 & 13 15 49.33  & +42 01 45.40  & 8.3 & I \\
  \enddata
 \label{tbl:dist}
 \end{deluxetable}


As discussed in \citet{thi07}, there are two types of XUV disks.  Type 1 disks have an extended region of UV-bright and optically-faint complexes that lay beyond the classical threshold for star formation in low-density environments.  These disks are represented here by NGC 0628, NGC 2841, NGC 3621, and NGC 5055.  NGC 2090 is a Type 2 disk, which is defined as having an extensive outer region within the SF threshold with optically low surface brightness and a blue (FUV-NIR) color.

Table~\ref{tbl:target} contains the locations of the target regions (also displayed in Figure~\ref{fig:whole}), as well as galaxy-specific exposure times and sensitivities. For the GALEX FUV and NUV filters \citep{mor07}, the exposure times range from 1644 s for NGC 0628 to 7269 s for NGC 3621, as compared to $\sim$1500 s integrations for NGS.  These correspond to a 3 sigma FUV sensitivity, corrected for Galactic extinction, of 3.46$\e{-18}$ and 2.0$\e{-18}$ erg s$^{-1}$ cm$^{-2}$ $\AA$$^{-1}$, respectively. 


\begin{deluxetable}{ccccccccc}
  \tabletypesize{\scriptsize}
  \tablecolumns{9}
  \tablecaption{Image Information}
  \tablewidth{0pt}
  \tablehead{
  \colhead{Target} &
  \colhead{RA} &
  \colhead{Dec} &
  \colhead{Galactocentric} &
  \multicolumn{3}{c}{Exposure Time (s)} &
  \multicolumn{2}{c}{3$\sigma$ Sensitivity\tablenotemark{a}} \\
  \noalign{\smallskip}
  \colhead{Field} &
  \colhead{(J2000)} &
  \colhead{(J2000)} &
  \colhead{Distance} &
  \colhead{FUV} &
  \colhead{NUV} &
  \colhead{IRAC} &
  \colhead{FUV} &
  \colhead{3.6 $\mu$m} 
  }
  \startdata
  NGC 0628-OUT & 01 37 02.00 & +15 55 05.00 & 9.5' (1.8 $R_{opt}$) & 1644.0 & 1644.0 & 1900.0 & 3.46$\e{-18}$ & 1.02$\e{-20}$ \\
  NGC 2090-OUT & 05 47 09.50 & $-$34 11 27.00 & 4.1' (1.7 $R_{opt}$) & 2281.0 & 4107.1 & 1800.0 & 3.95$\e{-18}$ & 9.60$\e{-21}$ \\
  NGC 2841-OUT & 09 22 21.50 & +50 50 33.00 & 9.3' (2.3 $R_{opt}$) & 1775.4 & 1775.5 & 1900.0 & 1.94$\e{-18}$ & 7.01$\e{-21}$ \\
  NGC 3621-OUT & 11 17 59.42 & $-$32 39 42.60 & 10.1' (1.7 $R_{opt}$) & 7269.1 & 7269.1 & 1800.0 & 2.00$\e{-18}$ & 1.00$\e{-20}$ \\
  NGC 5055-OUT & 13 16 21.70 & +41 51 42.00 & 12.9' (2.1 $R_{opt}$) & 1660.0 & 1660.0 & 1900.0 & 1.76$\e{-18}$ & 8.09$\e{-21}$ \\
  \enddata
 \tablenotetext{a}{Sensitivities are given in units of erg s$^{-1}$ cm$^{-2}$ $\AA^{-1}$ pixel$^{-1}$. The FUV sensitivities have been corrected for Galactic extinction (Table~\ref{tbl:sources}).}
 \label{tbl:target}
 \end{deluxetable} 


For the infrared observations, the Spitzer IRAC instrument \citep{faz04} observed the outer disk fields as part of GO Program 30753 (PI: D. Calzetti).  The fields were observed using two separate Astronomical Observing Requests (AORs) that each consisted of nine or ten 100 second dithered integrations (30 seconds for NGC 3621).  The dithered imaging allows for accurate removal of cosmic rays and detector artifacts, while executing the two AORs separately enables asteroid removal and image redundancy.  Imaging data were initially processed by the S14.4 version of the Spitzer Science Center IRAC Basic Calibrated Data (BCD) pipeline.  Post-processing image refinement was performed using the Local Volume Legacy (LVL; \citet{dal09}) imaging pipeline, which includes artifact correction, mosaicing, and sky subtraction.  For this study, the IRAC images were resampled, conserving surface brightness, from an original pixel scale of 0.75'' to match the GALEX pixel scale of 1.5''.  All four IRAC images and the GALEX FUV images were aligned and registered to the appropriate GALEX NUV image.  The final IRAC mosaics have total integration times of $\ge$1800 s per pixel, yielding an average 3 sigma 3.6 $\mu$m sensitivity of 8.98$\e{-21}$ erg s$^{-1}$ cm$^{-2}$ $\AA$$^{-1}$.  Total integration times and sensitivities for each galaxy are provided in Table 2.  

\section{Association Identification and Selection}

Given the resolution of our images, the star forming associations we seek to identify will appear as UV-bright point sources.  Following the procedures of \citet{don08}, we identify all 2$\sigma$ sources in the FUV image using 
SExtractor (version 2.4.4, \citet{ber96}), discarding any sources outside the IRAC FOV. Due to the low count rate for GALEX, its background is best treated as a Poissonian distribution whereas the background for Spitzer IRAC images can be treated as Gaussian.  Because of this difference, we employ the IRAF routine `phot' for photometry and background estimation.  The flux in each band is measured in circular apertures for each association, subtracting the local background through iterative rejection using the median value for GALEX and mode for Spitzer images.  For efficiency's sake, a single aperture size is chosen for each galaxy (see Table~\ref{tbl:sources}) and applied to all associations within that galaxy.  Since all of our associations are point sources at the resolution of GALEX and IRAC, we can then apply an appropriate aperture correction.

All apertures are equal to or greater than 5 pixels (7.5''), so no aperture correction is needed for the GALEX fluxes (FUV FWHM $\sim$4.5'').  To find the aperture correction for the IRAC bands, we referred to the IRAC Handbook and the point source aperture corrections found there.  For our smallest apertures, this amounted to a $\sim$5$\%$ correction. Taking into account the crowding of the field, a sufficiently large annulus for each source was chosen to measure the local background. In some cases, the flux of the source in one or more bands was very faint and with background subtraction, these fluxes became negative.  In these cases, a 3$\sigma$ upper limit is adopted.  The fluxes measured in all six bands are used to derive each association's spectral energy distribution (SED).  Since to determine the age/mass of an association we will need to fit a model to its SED, and since we have only a few datapoints per association (less than six, as will be explained in Section 4), those associations that are missing the NUV or 3.6 $\mu$m point are discarded and those missing the 4.5 $\mu$m point are flagged.  Finally, the two GALEX bands are corrected for foreground Galactic extinction using the values in Table~\ref{tbl:sources}.  


\begin{deluxetable}{ccccc}
  \tabletypesize{\scriptsize}
  \tablecaption{Photometry}
  \tablewidth{0pt}
  \tablehead{
  \colhead{Galaxy} &
  \colhead{Aperture} &
  \colhead{Foreground Galactic} &
  \colhead{Total} &
  \colhead{Final} \\
  \noalign{\smallskip}
  \colhead{ } &
  \colhead{Diameter (arcseconds)} &
  \colhead{Extinction $E(B-V)$ $\tablenotemark{a}$} &
  \colhead{$\#$ Sources} &
  \colhead{$\#$ Sources}
  }
  \startdata
  NGC 0628 & 7.5 & 0.07 & 48 & 32 \\
  NGC 2090 & 7.5 & 0.04 & 63 & 47 \\
  NGC 2841 & 10.5 & 0.016 & 87 & 75 \\
  NGC 3621 & 7.5 & 0.08 & 67 & 58 \\
  NGC 5055 & 12.0 & 0.018 & 146 & 17 \\
  \enddata
  \tablenotetext{a}{Extinction is applied as I$_{obs}$ = I$_{int}$$\times$10$^{-0.4K(\lambda)E(B-V)}$ where $K(\lambda)$ = 8.14 for FUV and 8.74 for NUV \citep{wyd05}.  Extinction values for each galaxy are from the dust maps of \citet{sch98}.}
 \label{tbl:sources}
 \end{deluxetable}

Two main sources of noise dominate the uncertainty on each flux measurement:  the noise from the source detection and the variance of the local background.  The noise from the source for UV and MIR bands is the Poissonian noise of the measurement, converted to electrons.  The background noise is the standard deviation on the sky pixels distribution, also converted into electrons.  This is then multiplied by the square root of the number of pixels in the aperture.  The two contributions are added in quadrature.  For IRAC data, the uncertainty is dominated by the variance in the background; for the UV bands, the dominant source of error is the uncertainty in the source counts.


The total number of sources is listed in Table~\ref{tbl:sources} for each galaxy.  Contaminating stars and galaxies are then eliminated with a number of methods.  From a visual inspection, foreground stars are removed by discarding any source with diffraction spikes in the original 3.6$\mu$m image.   Similarly, any source with visible extended emission at 3.6$\mu$m is rejected.  Dim, low-mass stars are eliminated by imposing the criteria: 


\begin{equation}\label{ratio}
\log \left(\frac{f_{\lambda}(FUV)}{f_{\lambda}(NUV)}\right) > -0.5
\end{equation}

which will eliminate sources that are red in the GALEX bands (lifetimes $\gtrsim$1 Gyr).  This criterion will also remove clusters with ages $\gtrsim$1 Gyr, but this is acceptable as our primary goal is to identify clusters of recent star formation.

\begin{deluxetable}{cllrrrrrr}
  \tabletypesize{\tiny}
  \tablecaption{Source List}
  \tablewidth{0pt}
  \tablehead{
   \colhead{Association} &
   \colhead{RA(J2000)} &
   \colhead{Dec(J2000)} &
   \colhead{FUV$\tablenotemark{a}$} &
   \colhead{NUV$\tablenotemark{a}$} &
   \colhead{3.6$\tablenotemark{a}$} &
   \colhead{4.5$\tablenotemark{a}$} &
   \colhead{5.8$\tablenotemark{a}$} &
   \colhead{8.0$\tablenotemark{a}$} \\
   \noalign{\smallskip}
   \colhead{ID} &
   \colhead{degree} &
   \colhead{degree} &
   \colhead{$F_{\lambda}$ (1$\sigma$)} &
   \colhead{$F_{\lambda}$ (1$\sigma$)} &
   \colhead{$F_{\lambda}$ (1$\sigma$)} &
   \colhead{$F_{\lambda}$ (1$\sigma$)} &
   \colhead{$F_{\lambda}$ (1$\sigma$)} &
   \colhead{$F_{\lambda}$ (1$\sigma$)} 
}
  \startdata
\multicolumn{9}{c}{} \\
\multicolumn{9}{c}{\underline{NGC 0628 Associations}} \nl \multicolumn{9}{c}{}\\ 
1 & 01 37 00.55 & +15 51 18.9 & 2437.54(194.23) & 900.84(51.99) & 3.09(0.47) & $-$0.58(0.94) & $-$8.84(1.25) & 4.04(0.50) \\
2 & 01 37 03.67 & +15 54 29.4 & 919.19(119.96) & 240.34(32.49) & 4.14(0.17) & 2.09(0.17) & $-$4.47(0.45) & 3.97(0.24) \\
3 & 01 37 04.82 & +15 54 27.9 & 814.75(113.36) & 296.35(32.88) & 14.79(0.18) & 8.58(0.16) & $-$6.37(0.45) & 6.18(0.26) \\
4 & 01 37 15.63 & +15 54 33.8 & 239.11(67.21) & 31.49(24.33) & 0.46(0.36) & 0.40(0.343) & 5.79(2.65) & 5.33(0.48) \\
5 & 01 36 56.92 & +15 54 11.5 & 283.53(73.81) & 230.79(31.92) & 0.42(0.18) & 0.30(0.19) & 6.88(1.53) & $-$4.58(0.85) \\
6 & 01 36 56.08 & +15 53 56.3 & 457.59(93.03) & 152.05(32.41) & 2.71(0.18) & 1.12(0.21) & $-$7.75(0.47) & $-$1.173(0.88) \\
\enddata
\tablecomments{(a) Fluxes are in units of
10$^{-19}$~ergs~s$^{-1}$~cm$^{-2}$~A$^{-1}$; the 1~$\sigma$ uncertainties
are indicated in parenthesis, following the flux value. For FUV
and NUV band, the flux has been corrected for the foreground Milky
Way extinction.}
\label{tbl:flux}
\end{deluxetable}

\clearpage

To quantify the remaining background galaxy contamination, we derived the number of potential contaminants within the region of interest ($\sim$1 square arcminute).  Our deepest observations are of NGC 3621, with a 3 sigma FUV detection limit corresponding to an AB magnitude of 25.9.  Taking into account that objects larger than 2'' will be resolved by IRAC and recognizable as extended sources, we can estimate the number of point sources within 1 square degree to be $\sim$$10^4$ \citep{xu05}.  This leaves us with a conservative 3$-$5 galaxies per square arcminute for NGC 3621, and less for our other shallower galaxies.  In reality, the number of contaminating galaxies is even less than our already conservative estimate.  Most galaxies are redder on average, due to older stellar populations or larger dust content, than the associations of interest here.  These galaxies are automatically discarded through Equation (1) because of their faint FUV emission.  Furthermore, dusty low redshift galaxies would present excess 8.0 $\mu$m emission from Polycyclic Aromatic Hydrocarbons (PAH) in the IRAC4 band (e.g. \citep{leg84, smi07}) which we do not detect in our images (see Section 4.1).  This analysis indicates that the level of background galaxy contamination in our sample is low and we neglect it in what follows. 

\begin{figure}[H]
\centering
\epsscale{1}\plotone{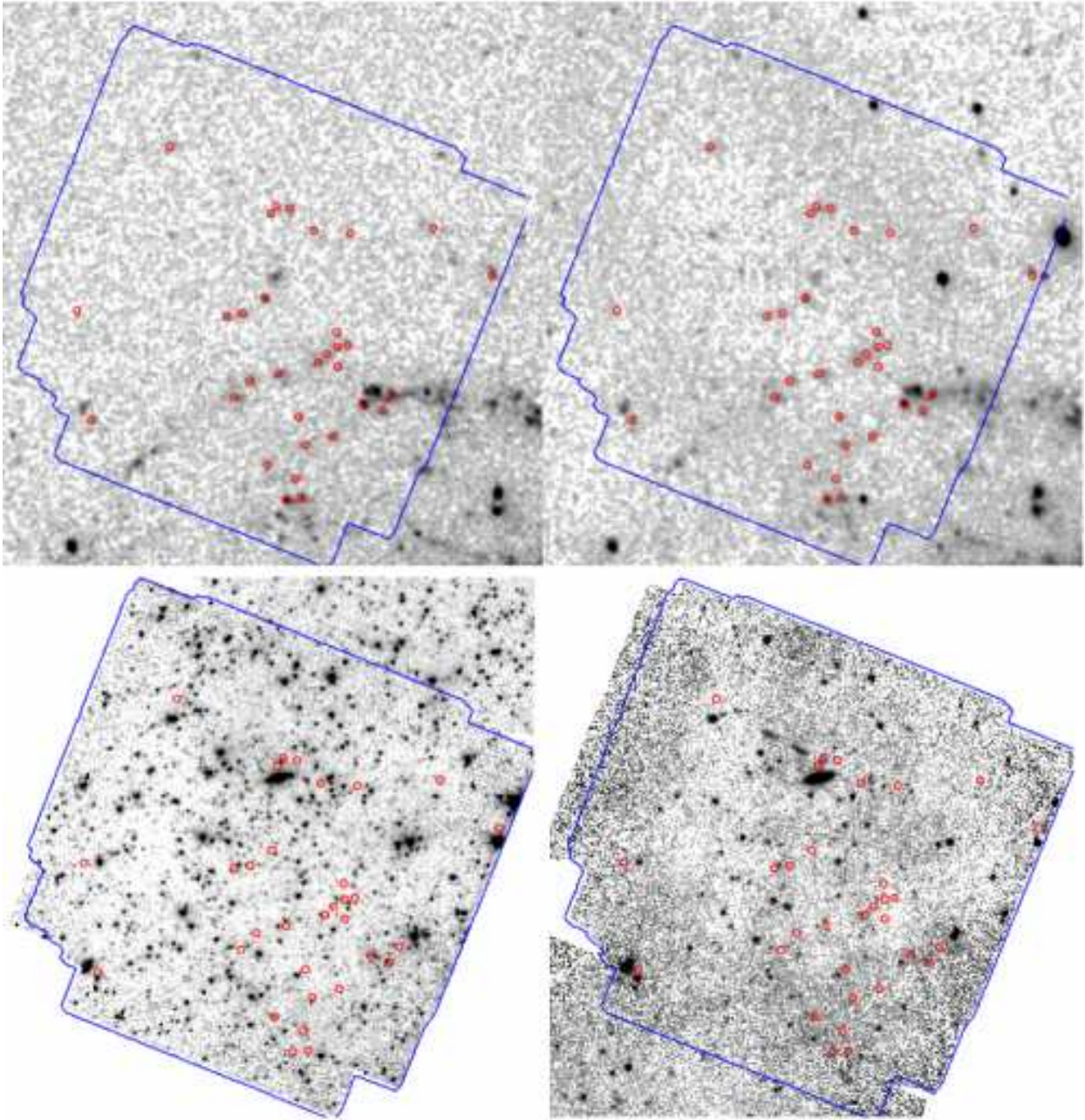} 
\caption{GALEX FUV (top left) and NUV (top right) and Spitzer 3.6 $\mu$m (bottom left) and 8.0 $\mu$m (bottom right) images of the target region in NGC 0628.  There are 32 UV-bright star forming associations which are indicated by the red apertures used in measuring the photometry (see Table \ref{tbl:sources} for aperture sizes).  UV sources that are not selected are either below our detection threshold, discarded due to missing the 3.6 $\mu$m band, or candidates for foreground stars or background galaxies.}
\label{fig:0628}
\end{figure}

\begin{figure}[H]
\centering
\epsscale{1}\plotone{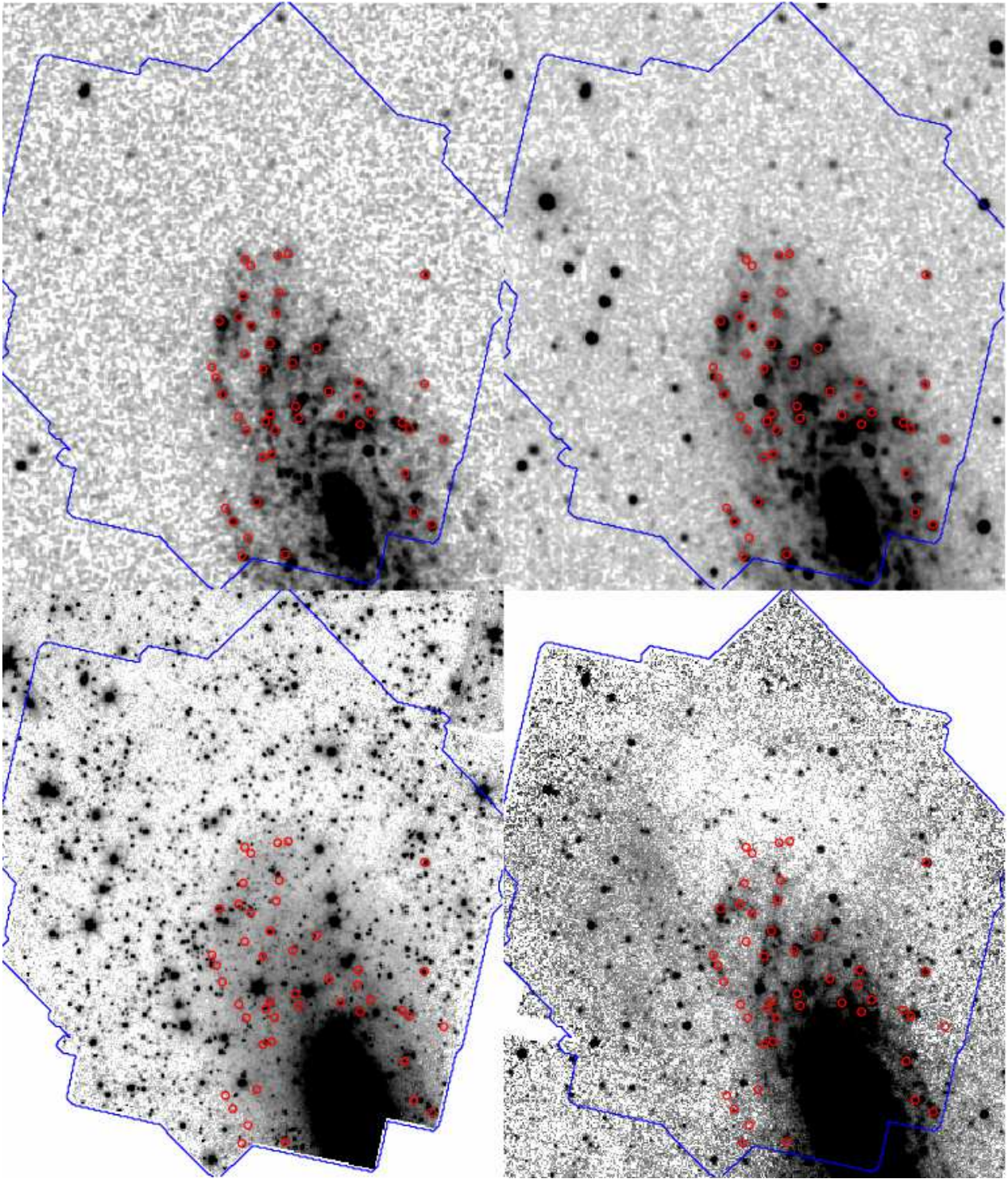} 
\caption{As in Figure~\ref{fig:0628}, for NGC 2090.  There are 47 final UV-bright associations.}
\label{fig:2090}
\end{figure}

\begin{figure}[H]
\centering
\epsscale{1}\plotone{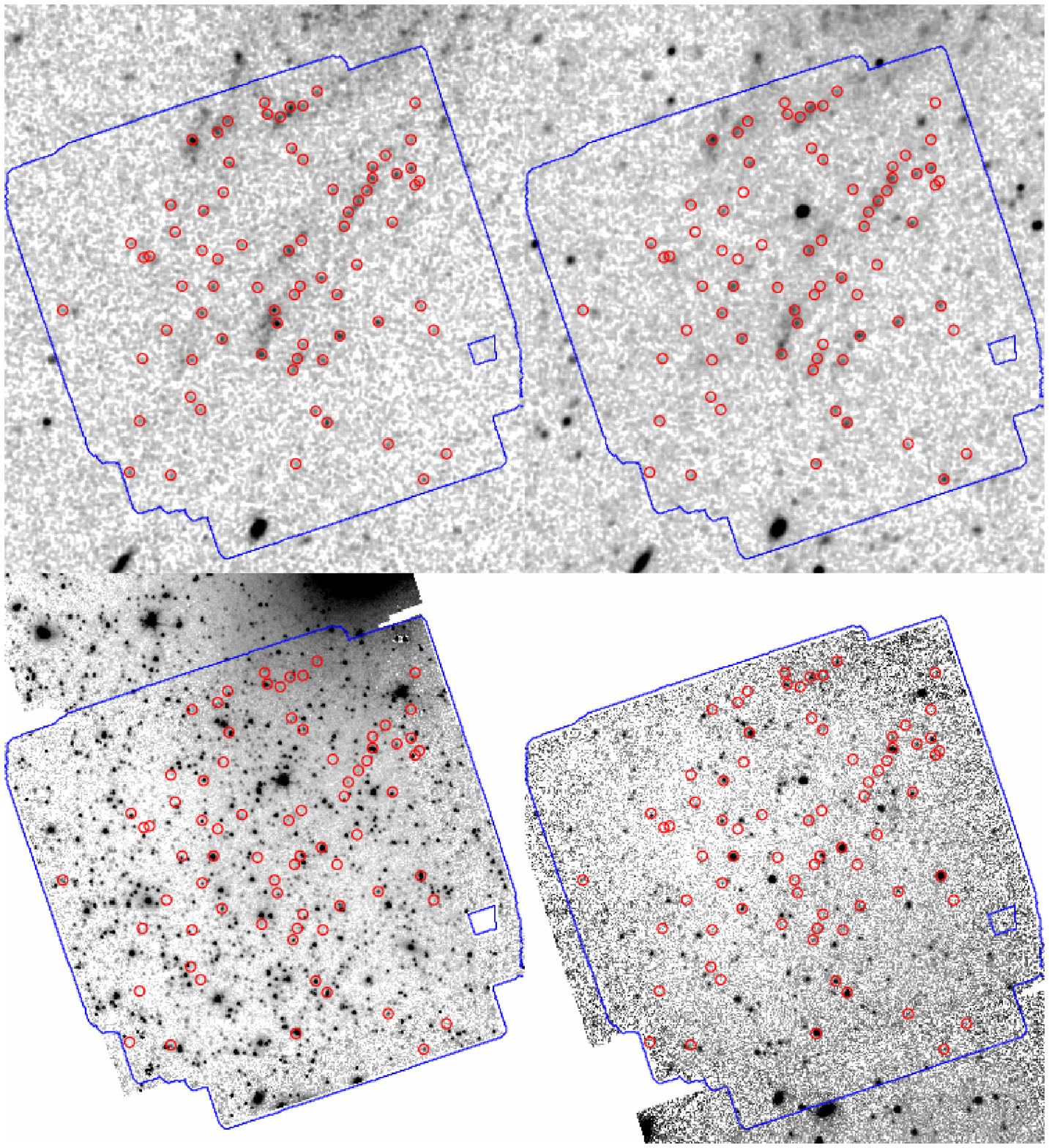}
\caption{As in Figure~\ref{fig:0628}, for NGC 2841.  There are 75 final UV-bright associations.}
\label{fig:2841}
\end{figure}

\begin{figure}[H]
\centering
\epsscale{1}\plotone{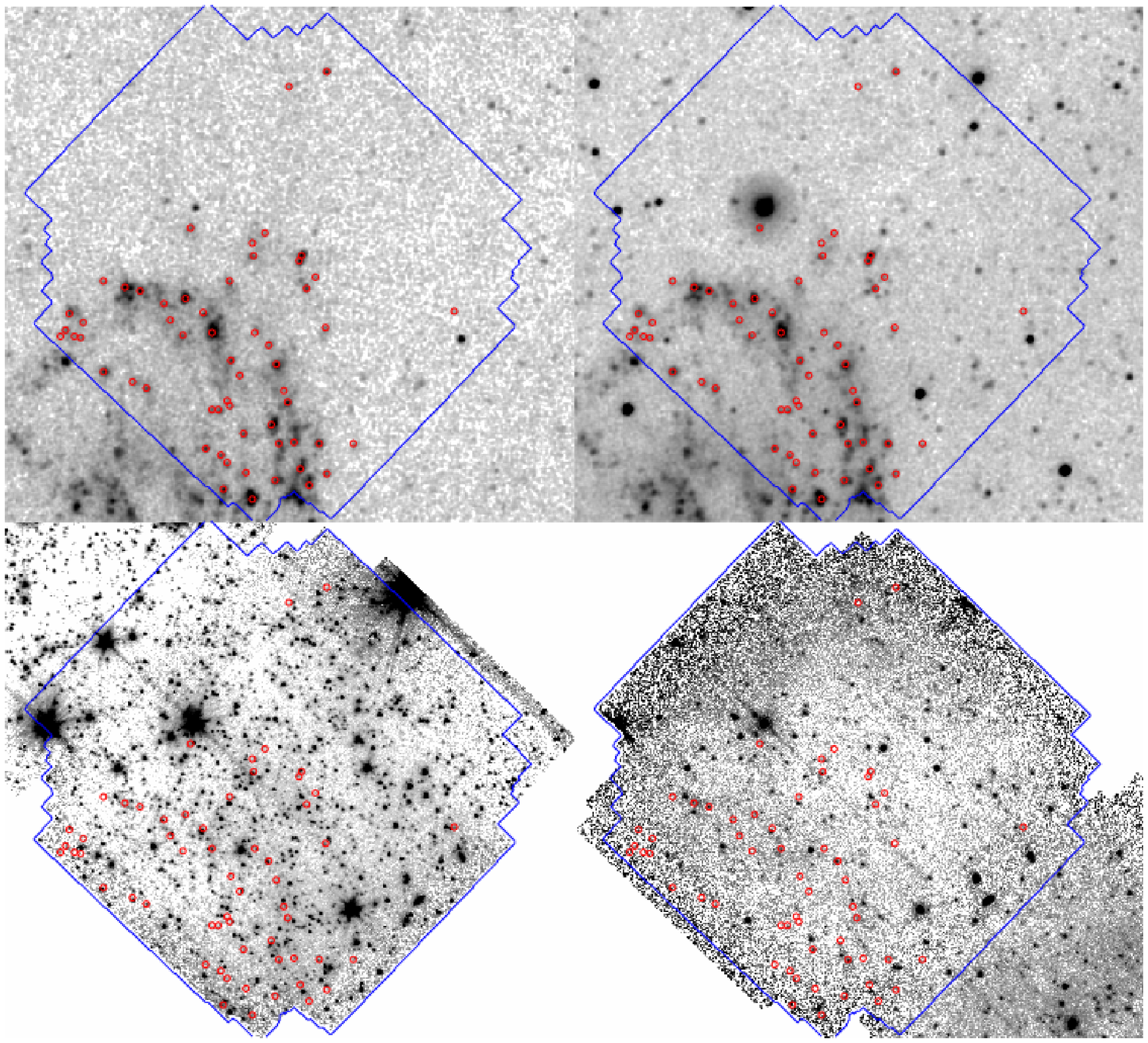}
\caption{As in Figure~\ref{fig:0628}, for NGC 3621.  There are 57 final UV-bright associations.}
\label{fig:3621}
\end{figure}

\begin{figure}[H]
\centering
\epsscale{1}\plotone{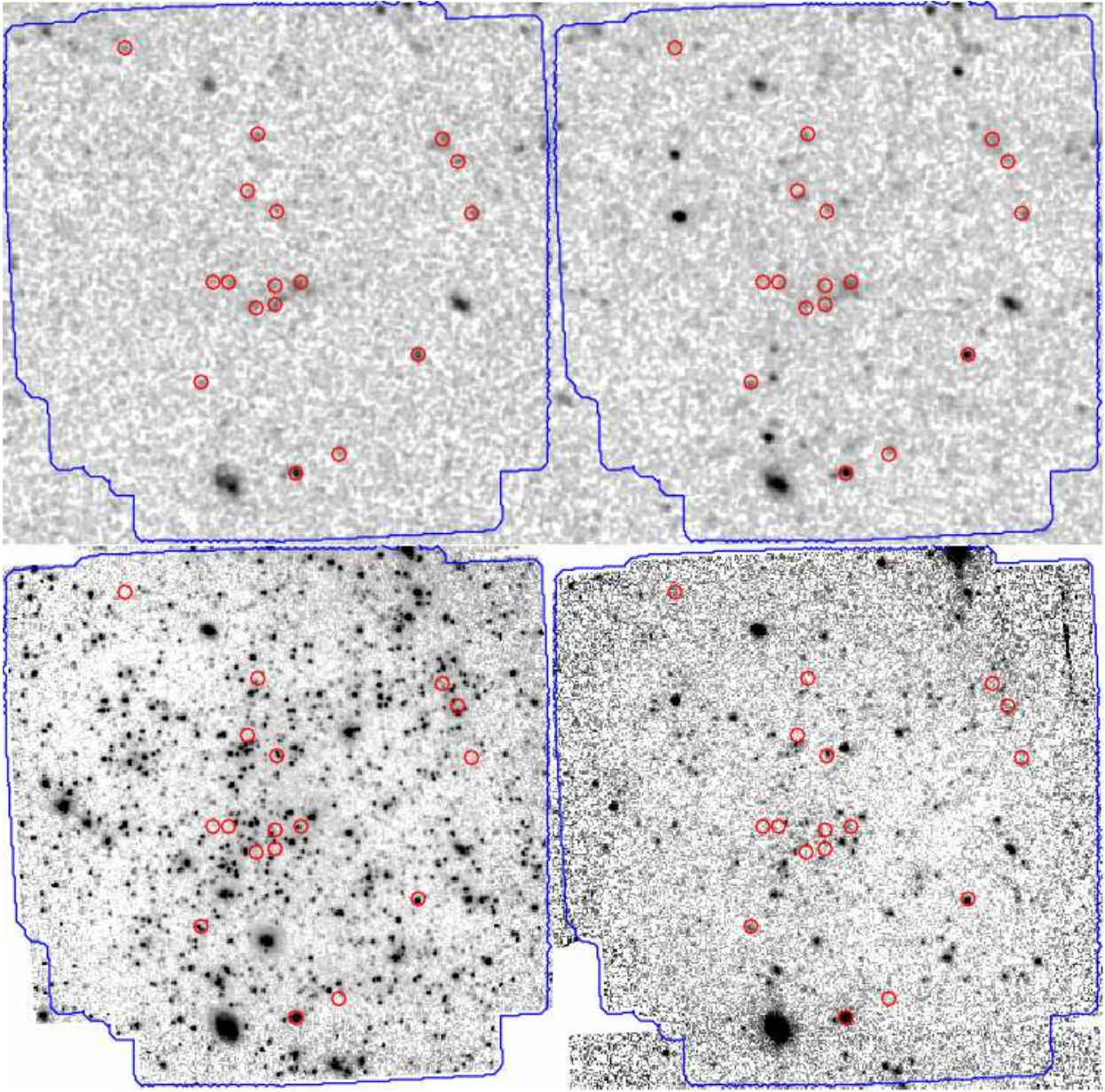}
\caption{As in Figure~\ref{fig:0628}, for NGC 5055.  There are only 17 final associations, due to intrinsically high noise levels (see Section 5.5).}
\label{fig:5055}
\end{figure}

Our final number of sources is 229.  Their locations, fluxes, and uncertainties are listed in Table~\ref{tbl:flux}.  Figures~\ref{fig:0628}$-$\ref{fig:5055} show the star forming associations on the FUV, NUV, 3.6 $\mu$m and 8.0 $\mu$m images of each galaxy.



\section{Analysis}

\subsection{Stellar Population Modeling}

To constrain the age and mass of each source we fit their UV+MIR SED with a model SED of a single age stellar population.  These model SEDs are a combination of nebular and stellar continuum emission and, as such, do not account for the effects of extinction, i.e. the redistribution of ultraviolet light into the infrared by dust.  We can make the reasonable assumption that the 3.6 $\mu$m and 4.5 $\mu$m bands are free of this effect \citep{pah04,cal05, eng08}.  This is not in general true for the 5.8 $\mu$m and 8.0 $\mu$m bands as they may contain a significant amount of dust emission including that of polycyclic aromatic hydrocarbons (PAHs).  As such, they will not be directly included in our fits, leaving us with four bands.  

To further address the question of dust, we note that \citet{gil07a} found sub-solar metallicities for the HII regions in the outer disk of M83 with Z/Z$_{\odot}$ = 1/5$-$1/10.  Coupled with the result that low metallicity regions are expected to also have under-luminous 8.0 $\mu$m dust emission \citep{eng05,eng08,cal07,dra07}, we infer that outer disks will have little dust in general.  To see if this is true for our galaxies and our particular associations, we create dust maps by taking the 8.0 $\mu$m image and subtracting off a scaled 3.6 $\mu$m image (which represents the continuum).  In general, these dust maps show weak or null dust detections near our associations, which is in agreement with the low metallicity values.  This will be discussed in more detail for each galaxy in Section 5.

All of our fits are done using Starburst99 stellar population synthesis models \citep{lei99,vaz05}.  Our associations are small (60$-$110 pc) and therefore most likely contain a single or a few stellar clusters, which we model as instantaneous burst populations with sub-solar metallicities (Z = 0.004).  Comparisons between sub-solar and solar metallicity models show only negligible differences at the wavelengths utilized in this study. Initially the IMF (see also Section 6) is chosen to be the default Kroupa IMF, which is expressed by a broken power law ($\phi$(M) $\propto$ M$^{-\alpha}$) with a slope of $\alpha$ = 1.3 for 0.1 $\Msun$ $<$ $M$ $<$ 0.5 $\Msun$ and $\alpha$ = 2.3 for 0.5 $\Msun$ $<$ $M$ $<$ 100 $\Msun$ \citep{kro01}.  We employ the Padova tracks with full asymptotic giant branch (AGB) evolution, the latter to better fit the MIR fluxes.  The contribution by thermally-pulsating AGB stars should be minimal for associations much younger than 1 Gyr \citep{mar06}.  Models are generated in timesteps of 1 Myr for 1$-$10 Myr and 5 Myr for 10 Myr$-$1 Gyr and then convolved with the GALEX and IRAC filter bandpasses to create the final synthetic photometry.

\subsection{Age and Mass Fitting}

\begin{figure}[H]
\epsscale{1.1}\plottwo{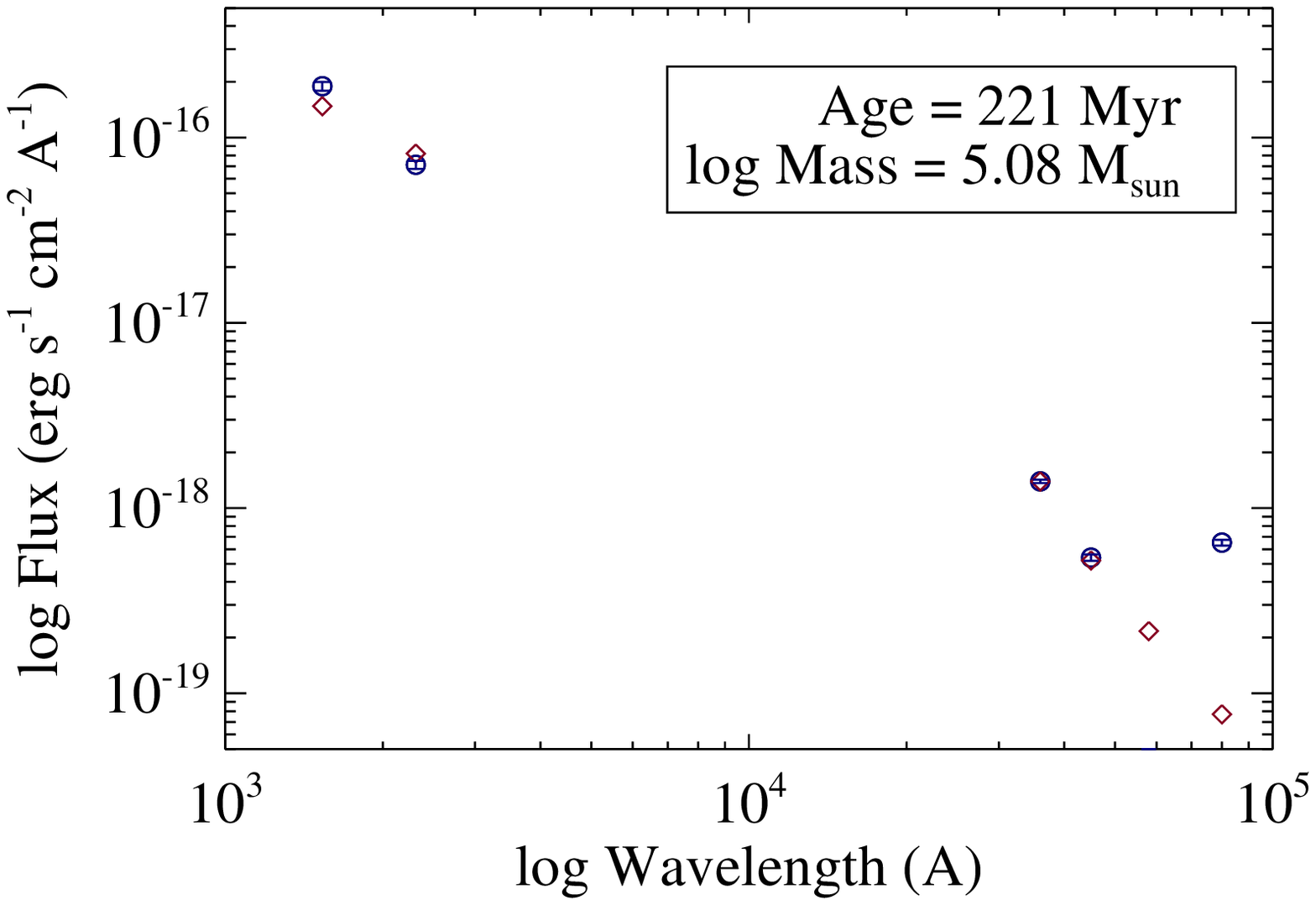}{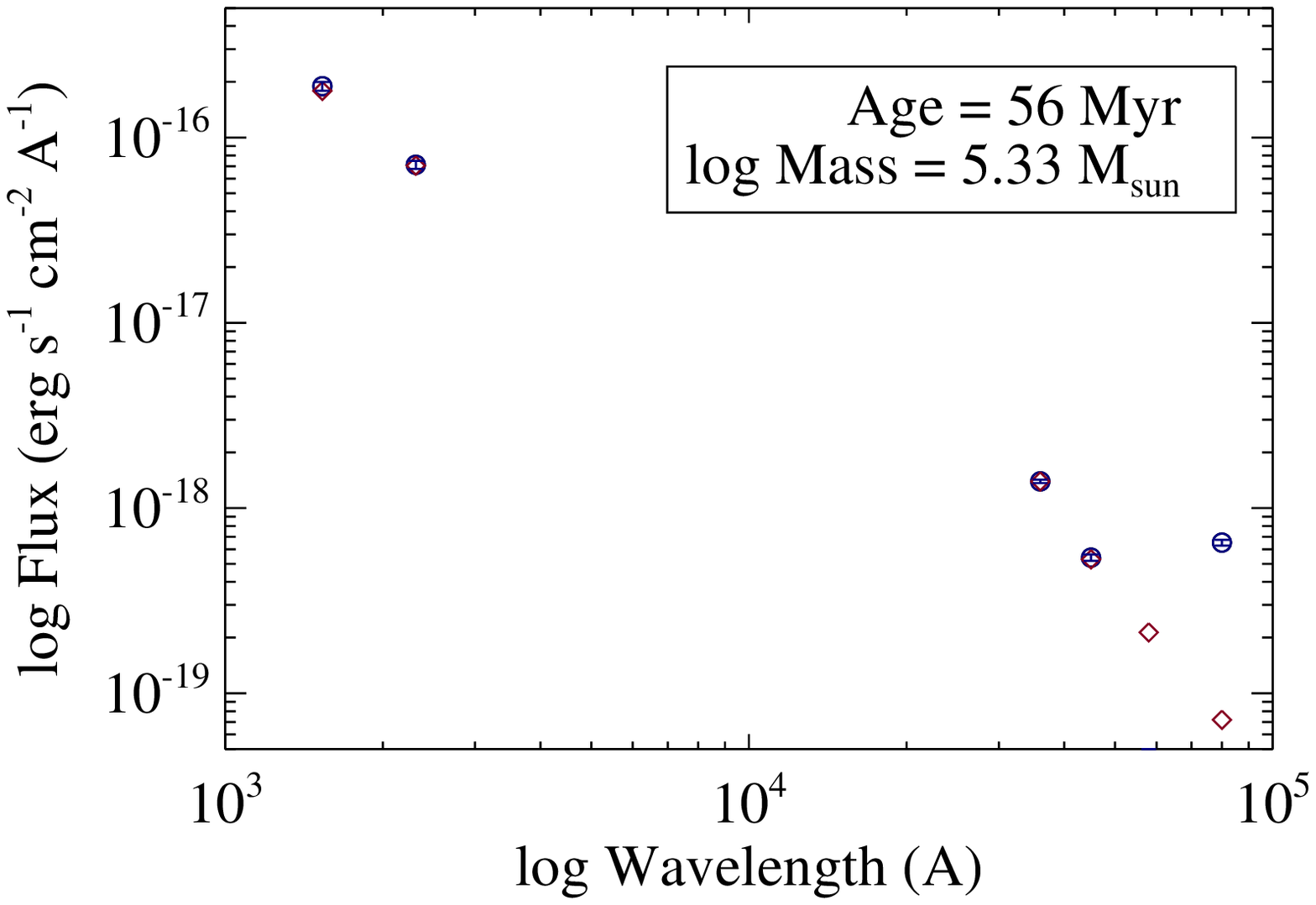}
\caption{Association 13 in NGC 3621 is a good example of an association that is well fit in both the no dust (left) and $E(B-V)$ = 0.3 (right) cases.  The circles (blue) with error bars represent the datapoints from this association and the diamonds (red) represent the model points.  The 5.8 and 8.0 $\mu$m points are not used in any fits due to PAH emission contamination.  In this particular case, the 5.8 $\mu$m emission is not detected.}
\label{fig:fits}
\end{figure}

Observational data are fit with the synthetic photometry from models using $\chi^2$-minimization \citep{don08}:  

\begin{equation}\label{min}
\chi^2(t, E(B-V), M, Z) = \sum_N \frac{\left(L_{obs} - A \times L_{model}\right)^2}{\sigma_{obs}^2}
\end{equation}

where N is the number of filters per source,  $L_{obs}$ is the observed luminosity, $L_{model}$ is the model luminosity, and $\sigma_{obs}$ is the measurement uncertainty.  Effectively the age of each association is found by comparing the $shape$ of each association's SED to the $shape$ of each model, i.e. the intensity in UV relative to the intensity in the IR bands.  The mass is an $a$ $priori$ unknown scaling factor which is represented by the dimensionless parameter $A$, where $A$ = $\frac{M}{10^6 \Msun}$, $M$ is the mass in the aperture, and 10$^6$ $\Msun$ is the default mass of the Starburst99 models.  The underlying stellar components (and thus the mass) of an association are best traced by the 3.6 $\mu$m emission and so we use this as our guideline to find the mass scaling factor; however, we do not fix the mass to this point, but allow the scaling factor to vary within the 3.6 $\mu$m 1$\sigma$ errors during fitting.  In the few cases of very large errors, we also impose a minimum detectable mass derived by combining the FUV 3$\sigma$ detection threshold with a 1~Myr stellar population model, our bluest synthetic SED.  A good fit (within the 1$\sigma$ errors) is shown in Figure~\ref{fig:fits}.

The age-extinction degeneracy prevents us from fitting for age and extinction simultaneously.  Though we have reason to believe (as outlined above) that these regions are low-dust and low-metallicity, we generate fits for two fixed extinctions (in addition to foreground Galactic extinction, hereafter implied): $E(B-V)$ = 0 and $E(B-V)$ = 0.3, the latter being the average extinction value for inner disks and the high end of the extinction values observed in the outer disk of M83 \citep{gil07a}.  Extinction is applied using Milky Way-type dust with $A_{FUV}$/$E(B-V)$ = 8.38 and $A_{NUV}$/$E(B-V)$ = 8.74 \citep{wyd05}.  The actual extinction is specific to a given association and difficult to ascertain with the available information.  However, by analyzing both extreme fits, we can make determinations about the relative extinction of these galaxies as a whole and make predictions about the `best fit' for a given association within a set of reasonable solutions.  In general, we also note that higher extinction values are found for younger associations and lower extinctions for older associations.  We also fit the associations in our most extincted galaxy (NGC 2841, see Section 5.4) using the starburst attenuation curve of \citet{cal00}, but find that the results are similar to those obtained using the Milky Way extinction curve plus foreground dust.  We ascribe this to the small extinction values that characterize our regions.

\begin{figure}[H]
\epsscale{1.1}\plottwo{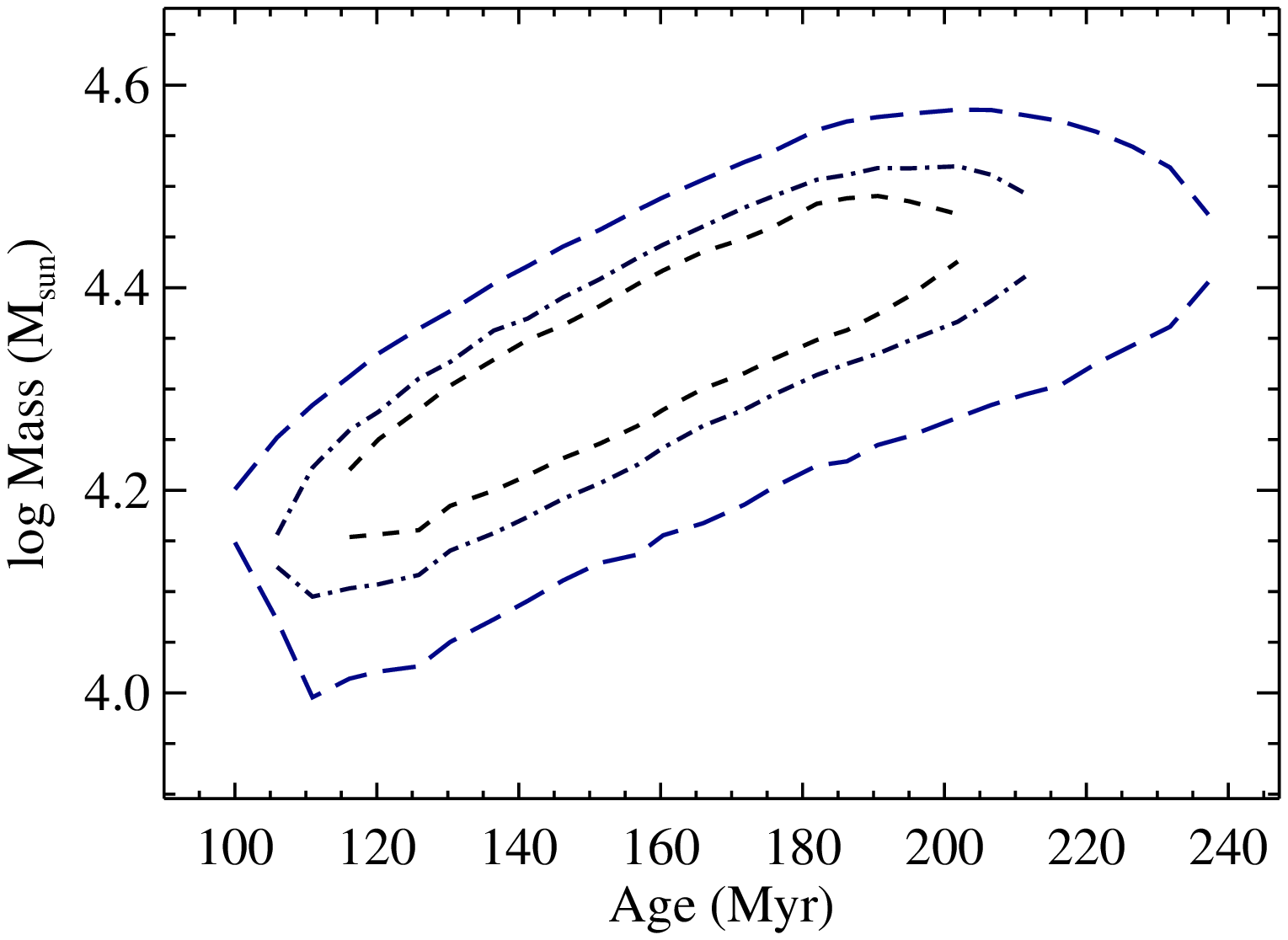}{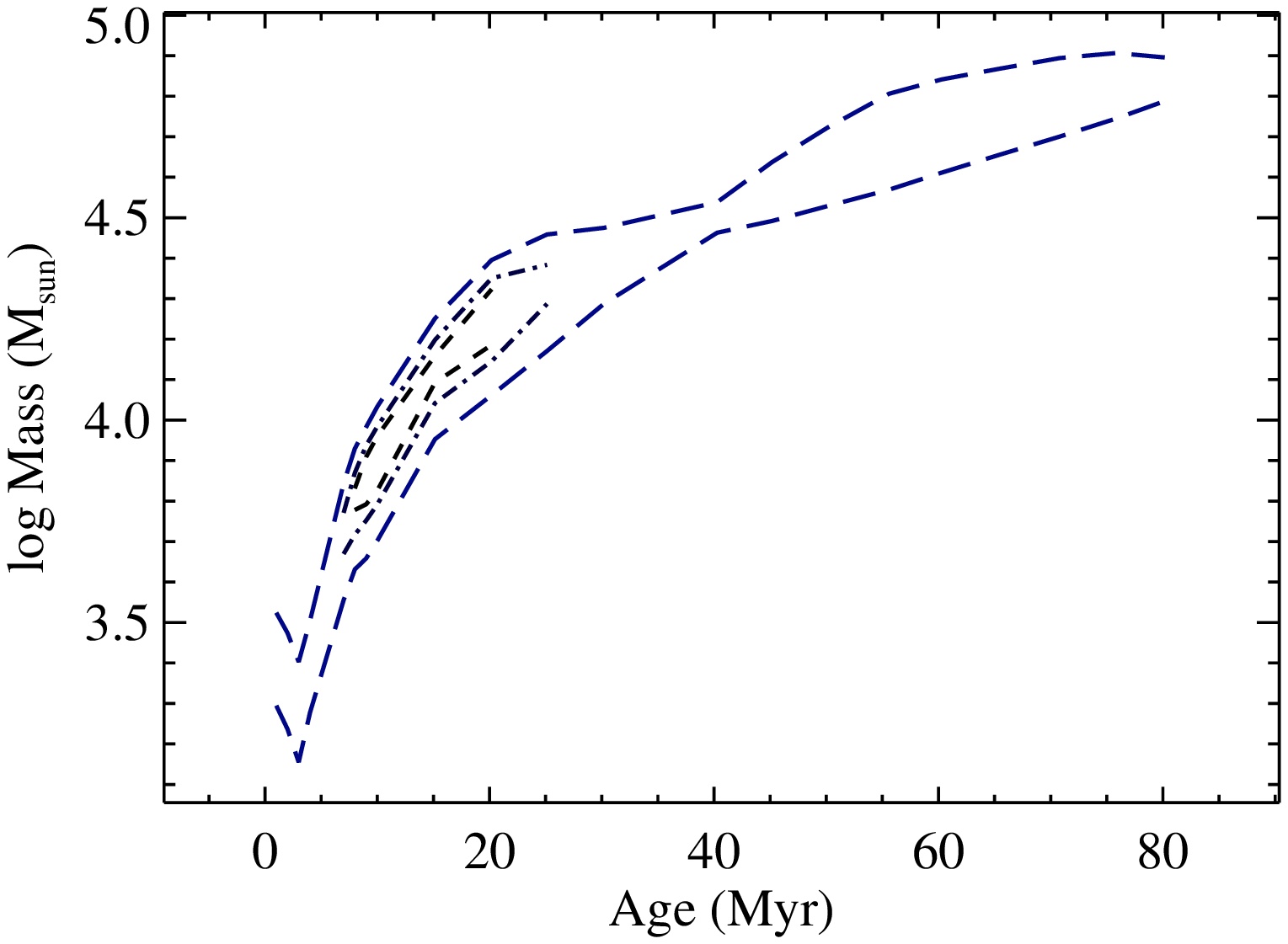}
\caption{Confidence contours for association 21 in NGC 3621 with $E(B-V)$ = 0.0 (left) and $E(B-V)$ = 0.3 (right).  This association's best fit age is $\sim$166 Myr and, in terms of  dust content, is not significantly different from the other associations in this or other galaxies.  The short dashed black line is at a 68$\%$ confidence level, the dot dashed purple line at 90$\%$, and the long dashed blue line is 99$\%$.}
\label{fig:con}
\end{figure}

The age and mass of each association are listed in Table~\ref{tbl:ages}, with associations separated for each galaxy into two groups:  those better fit by $E(B-V)$ = 0 and those better fit by $E(B-V)$ = 0.3.  The 1$\sigma$ uncertainties were determined by generating age/mass confidence intervals for each association.  An example of a typical confidence interval (at 68$\%$, 90$\%$, and 99$\%$ levels) can be seen in Figure~\ref{fig:con}.  It should be noted that, because the mass is tied to the value of the 3.6$\mu$m datapoint, increasing the mass will result in a decreased UV to MIR ratio, which corresponds to an older age. In other words, there is a positive correlation between age and mass.



\begin{deluxetable}{cllcc}
  \tabletypesize{\tiny}
  \tablecaption{Derived Ages and Masses for All Associations}
  \tablewidth{0pt}
  \tablehead{
  \colhead{Association} &
  \multicolumn{2}{c}{Age (Myr)} &
  \multicolumn{2}{c}{log Mass ($\Msun$)} \\
  \noalign{\smallskip}
  \colhead{ID } &
  \colhead{$E(B-V)$ = 0.0} &
  \colhead{$E(B-V)$ = 0.3} & 
  \colhead{$E(B-V)$ = 0.0} &
  \colhead{$E(B-V)$ = 0.3} 
  }
  \startdata
\multicolumn{5}{c}{} \\
\multicolumn{5}{c}{\underline{NGC 0628 Associations}} \\
\multicolumn{5}{c}{Best Fit with No Dust} \nl \multicolumn{3}{c}{} \\ 
3&$989^{+>11}_{-5}$&$116^{+5}_{-5}$&$5.82^{+0.01}_{-0.01}$&$5.28^{+0.03}_{-0.00}$ \\
5&$90^{+10}_{-40}$&$6^{+1}_{-1}$&$4.02^{+0.10}_{-0.10}$&$3.53^{+0.07}_{-0.17}$ \\
8&$120^{+16}_{-4}$&$15^{+5}_{-5}$&$4.46^{+0.04}_{-0.04}$&$4.39^{+0.03}_{-0.04}$ \\
12&$35^{+60}_{-5}$&$6^{+1}_{-1}$&$4.40^{+0.50}_{-0.50}$&$4.41^{+0.03}_{-0.02}$ \\
14&$106^{+5}_{-5}$&$7^{+1}_{-3}$&$4.57^{+0.03}_{-0.03}$&$4.10^{+0.03}_{-0.34}$ \\
\enddata
\tablenotetext{a}{These associations are missing their 4.5 $\mu$m point and are thus only fit with three datapoints.}
\tablenotetext{b}{These associations display an inverse correlation between age and mass for $E(B-V)$ = 0.}
 \label{tbl:ages}
 \end{deluxetable}

\clearpage

\section{Results}

\subsection{Dust vs. No Dust}

Our results are presented below for each galaxy in three categories:  `no dust' ($E(B-V)$ = 0), `dust' ($E(B-V)$ = 0.3), and `best fit' (either dust or no dust).  We use two criteria to determine the `best fit' category.  

\begin{enumerate}
\item As stated above, there exists a positive correlation between the age and mass of an association.  In 47 ($\sim$21$\%$) cases, however, we see an inverse correlation in the confidence intervals of cases involving no dust.  Figure~\ref{fig:con2} shows a typical example.  All of these cases become positively correlated when dust is added to the system and, for $\sim$68$\%$ of these cases, adding dust results in a lower minimum $\chi^2$ value by $\sim$49$\%$ on average.  Finally, $\sim$83$\%$ of these cases show excess 8.0 micron emission, indicating PAH emission in the vicinity of these associations.  Therefore, we conclude that these regions are good candidates for higher relative extinction and that the inverse correlation is an artifact of improperly treating the extinction in the no dust case.

\begin{figure}[H]
\epsscale{1.1}\plottwo{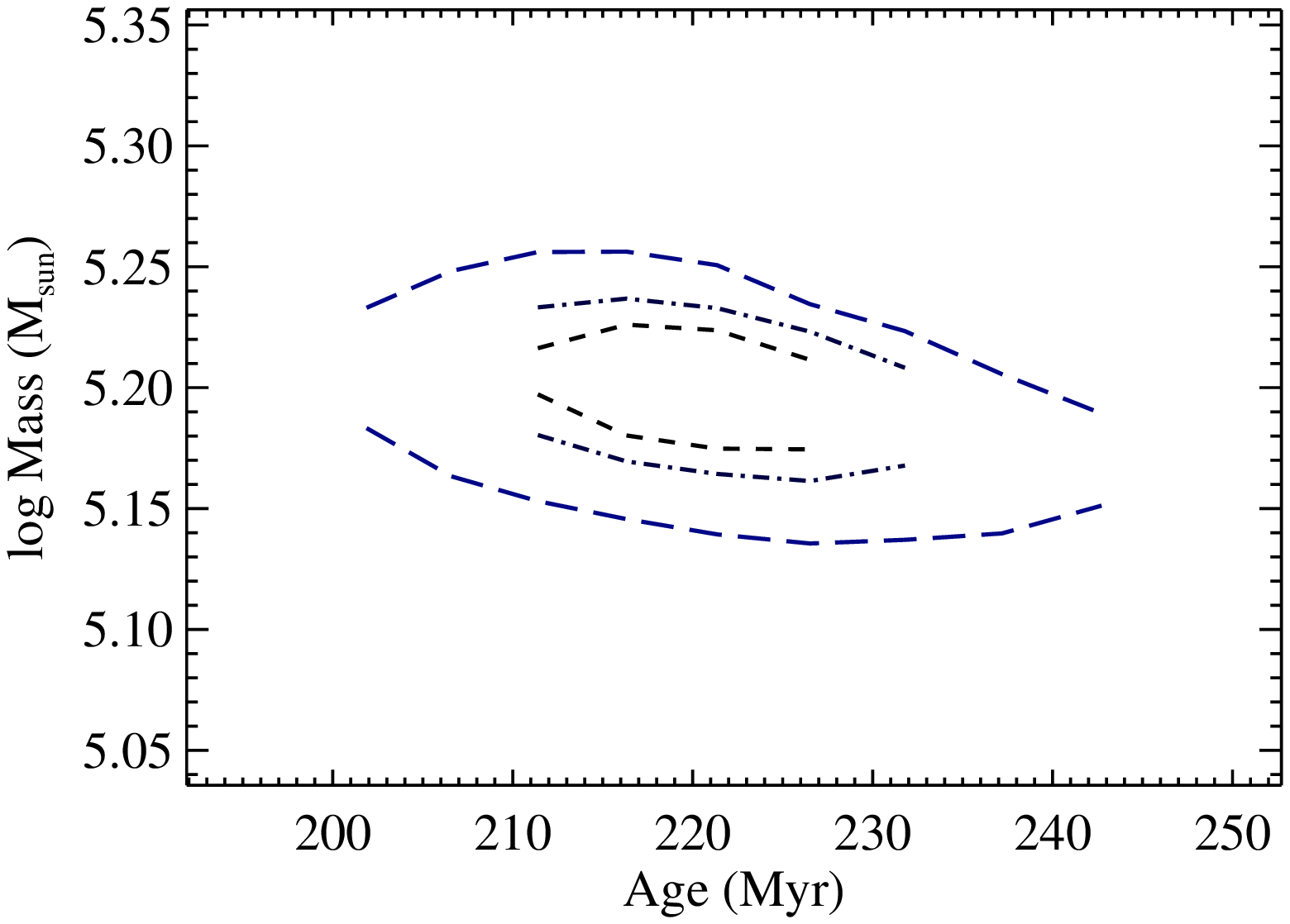}{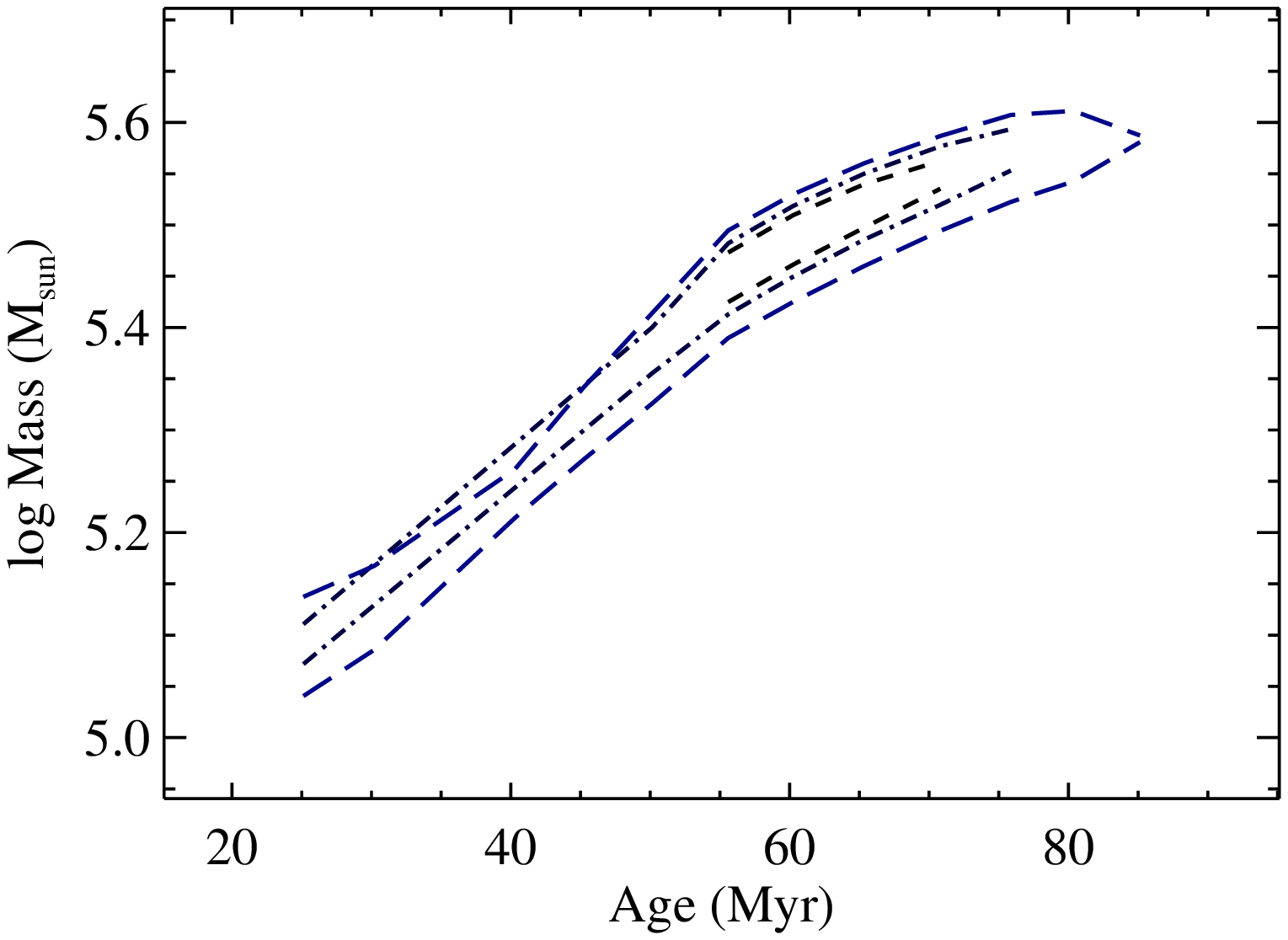}
\caption{Association 74 from NGC 2841 displays a typical inverse correlation (left) in the confidence intervals between age and mass.  When dust is added to the models, association 74 presents the expected positive correlation between age and mass (right).}
\label{fig:con2}
\end{figure}

\item For all remaining associations in which the positive correlation between age and mass appears for both the no dust and dust cases, the best fit is determined by the reduced $\chi^2$ value. $\sim$34$\%$ of associations have a lower reduced $\chi^2$ with no added dust and $\sim$45$\%$ of associations have a lower $\chi^2$ value when modeled with dust ($E(B-V)$ = 0.3). However, it should be noted that both fits are often acceptable (i.e. within the 1$\sigma$ errors) and so we report both solutions while counting these associations as dusty for the `best fit' statistics.    The distribution of minimum $\chi^2$ between the dust and no dust cases will be discussed on a galaxy-by-galaxy basis below. 

\end{enumerate}

Associations are further divided into three categories:  ionizing ($<$ 10 Myr), non-ionizing (10 Myr$-$1 Gyr), and evolved ($>$ 1 Gyr).  Since our criteria for removing contaminating foreground stars (e.g. Equation (1)) and background galaxies also deplete our sample of associations older than $\sim$1 Gyr, our statistics for such associations would be biased and we will not discuss the evolved associations any further.

All $\chi^2$ values shown or discussed are reduced with N = 3 and minimum $\chi^2$ comparisons are based on median values.

\subsection{NGC 0628}
   
\begin{figure}[H]
\epsscale{1}\plottwo{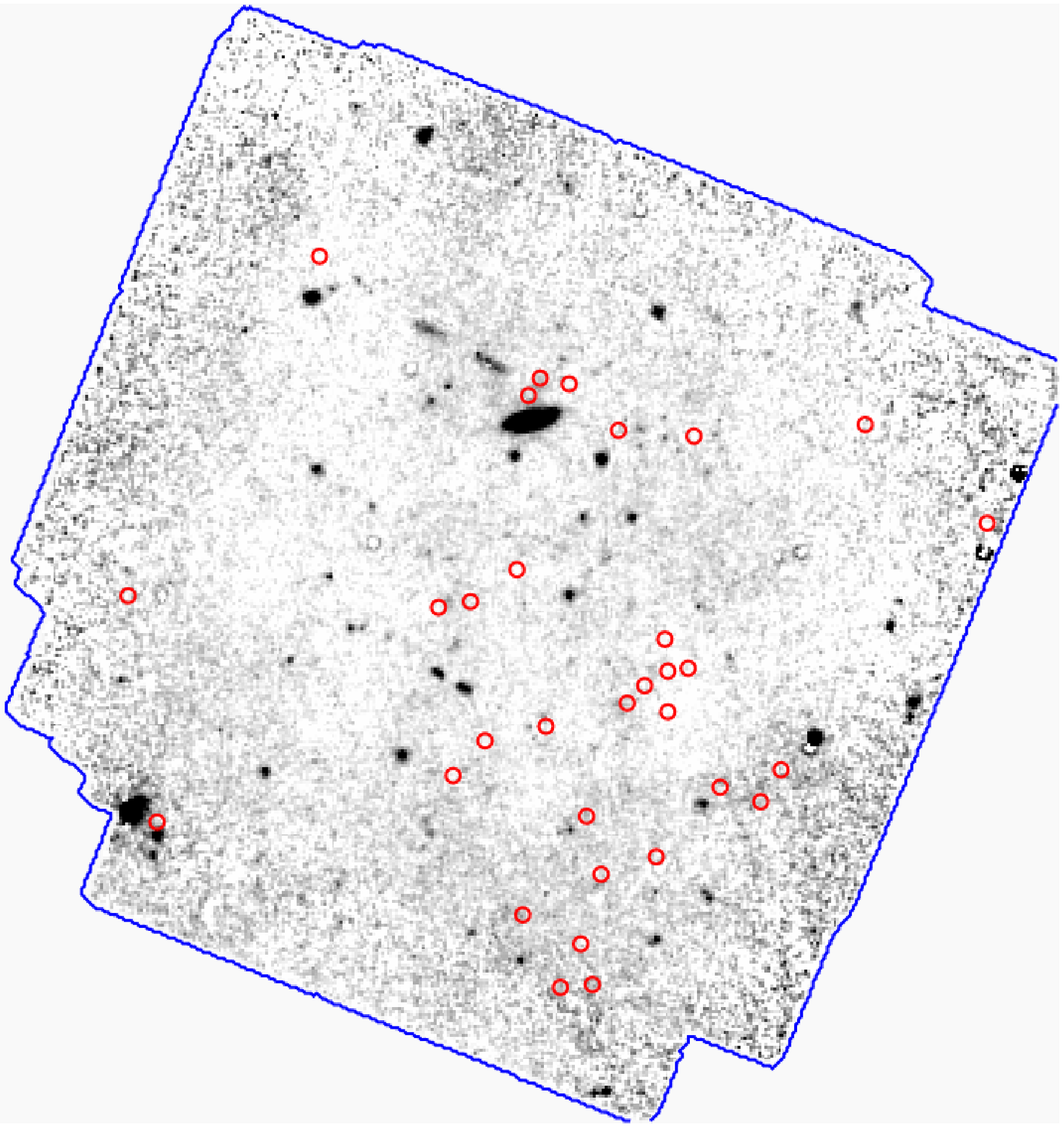}{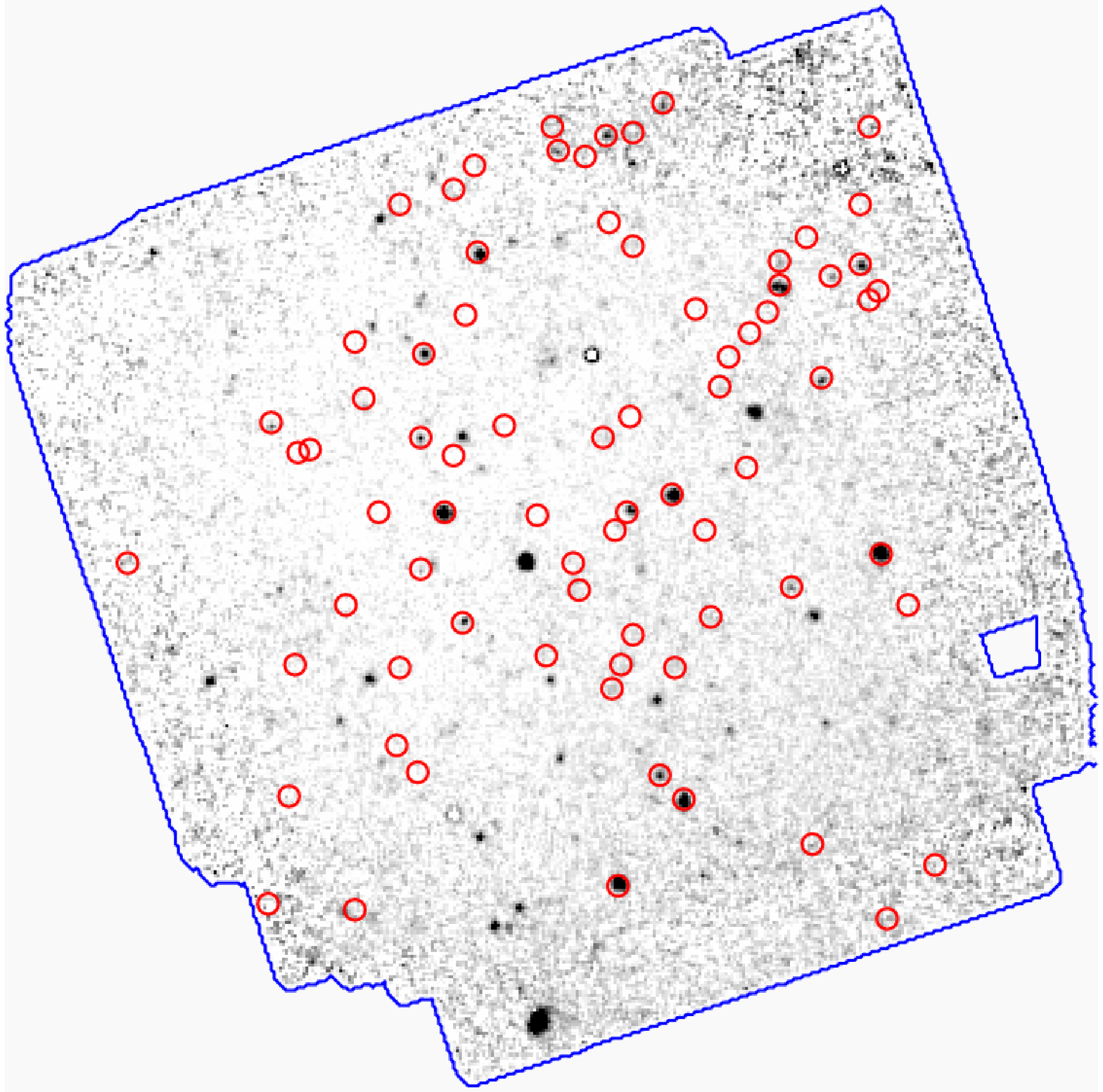}
\caption{The dust maps of NGC 0628 (left) and NGC 2841 (right) were created by scaling the 3.6 $\mu$m image to approximate the continuum emission and then subtracting from the 8.0 $\mu$m image.  Bright sources are indicative of dust excess as traced by the 8.0 $\mu$m (PAH emission), though these sources may be background galaxies due to their intrinsic dust emission.  The red apertures indicate UV bright stellar complexes.  It is clear that not many UV bright regions are associated with dust, though there are more in NGC 2841, which is supported by overall better $\chi^2$ values found when dust extinction is applied to the observed SEDs (Section 5.4).  It should be noted that the source of the dust emission and the source of the UV emission are not necessarily associated. }
\label{fig:dust}
\end{figure}

\begin{figure}[H]
\epsscale{0.7}\plotone{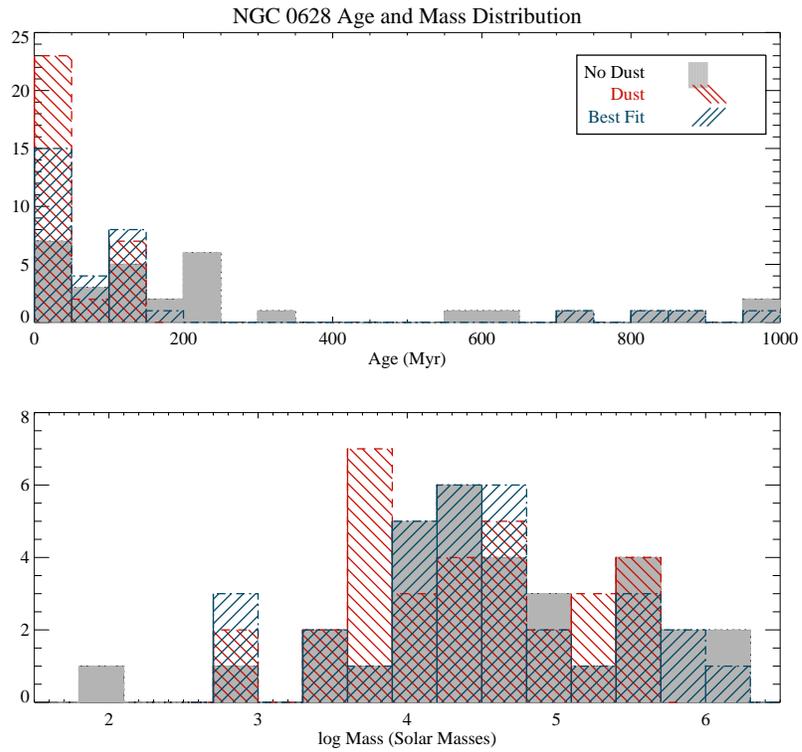}
\caption{The age and mass distribution for NGC 0628.  The no dust $E(B-V)$ = 0.0, dust $E(B-V)$ = 0.3, and the `best fit' case are shown, as indicated in the legend.}
\label{fig:0628age}
\end{figure}

\begin{figure}[H]
\epsscale{0.7}\plotone{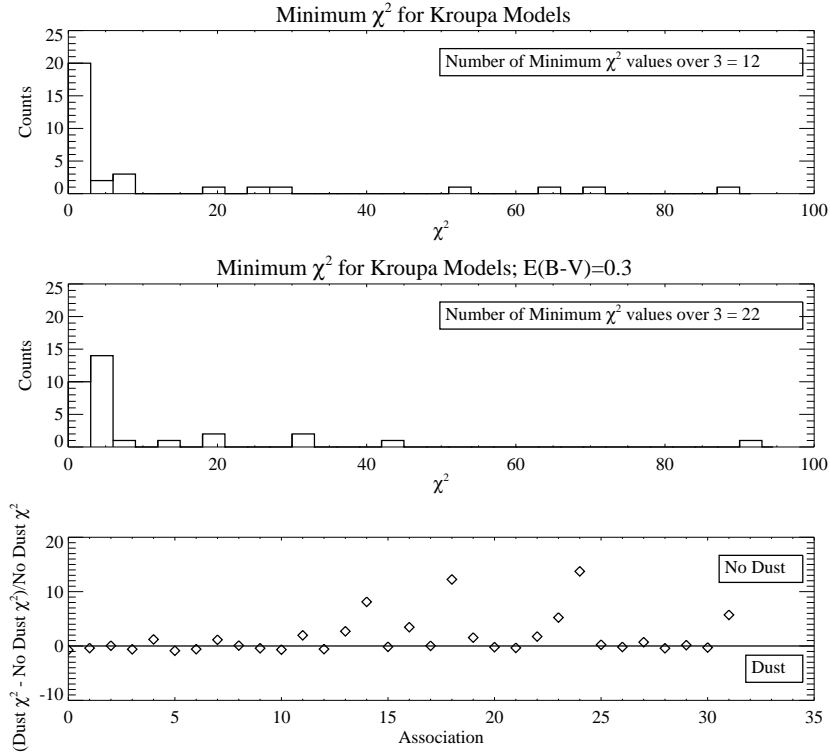}
\caption{NGC 0628 - The $\chi^2$ distribution demonstrates that overall this galaxy has a better fit without dust (top) than with it (middle).  Combined with the dust map and the number of occurrences of inverse correlation plots, this leads us to conclude that this is a relatively low-dust galaxy.  The bottom panel shows the relative degree of improvement in the $\chi^2$ value for the no dust or dust cases.  When analysed by association a few sources indicates a much better fit with dust.  Further studies are needed to ascertain the true nature of the dust on these individual scales.}
\label{fig:0628chi}
\end{figure}

The imaged region of NGC 0628 contains 32 star forming associations.  A dust map of this field (Figure~\ref{fig:dust} (left)) is consistent with low extinction in excess of the foreground as almost none of our stellar complexes are associated with dust emission.  Table \ref{tbl:0628} shows the break-down of associations into types, the average and median age and mass of all associations, and the age dispersion.  The latter spans the entire range of our models when no dust is applied.  When the most extreme dust case is applied to all associations, the age range is still large enough to account for a lack of H$\alpha$ relative to UV.  Figure~\ref{fig:0628age} shows the age and mass distribution.  Figure~\ref{fig:0628chi} shows the relative $\chi^2$ distribution association by association comparing the no dust case to the dust case.  The no dust case results in lower minimum $\chi^2$ values by 8$\%$. 7 associations ($\sim$22$\%$) show an inverse correlation between age and mass with no dust, with 5 ($\sim$16$\%$) showing both an inverse correlation and a better $\chi^2$ with dust.  8 ($\sim$25$\%$) additional associations have a better $\chi^2$ with dust.  These numbers are reflected in the best fit values.

\begin{deluxetable}{cc ccccc}
  \tabletypesize{\scriptsize}
  \tablecaption{NGC 0628 Association Statistics}
  \tablewidth{0pt}
  \tablehead{
  \colhead{ } &
  \colhead{Ionizing} &
  \colhead{Non-Ionizing} &
  \colhead{Evolved}  &
  \colhead{Average (Median)} &
  \colhead{log Average (Median)} &
  \colhead{Age}  \\
  \noalign{\smallskip}
  \colhead{ } &
  \colhead{($<$ 10 Myr) } &
  \colhead{(10 Myr$-$1 Gyr)} &
  \colhead{($>$ 1 Gyr) } &
  \colhead{Age [Myr]} &
  \colhead{Mass [$\Msun$]} &
  \colhead{Range [Myr]}
  }
  \startdata
  No Dust & 2 ($\sim$6$\%$)& 29 ($\sim$90$\%$) & 1 ($\sim$3$\%$) & 297 (191) &  4.69 (4.57)  & 1$-$1000 \\
  Dust & 16 ($\sim$50$\%$) & 16 ($\sim$50$\%$) & 0 & 37 (15) & 4.39 (4.39) & 1$-$125 \\
  Best Fit & 6 ($\sim$18$\%$) & 26 ($\sim$81$\%$) & 0 & 158 (60) & 4.48 (4.48) & 1$-$988
  \enddata
 \label{tbl:0628}
 \end{deluxetable}

With the best fit values for age, the percentage of ionizing associations is $\sim$18$\%$, which is consistent with H$\alpha$ observations in outer disks.


\subsection{NGC 2090}

\begin{figure}[H]
\epsscale{0.6}\plotone{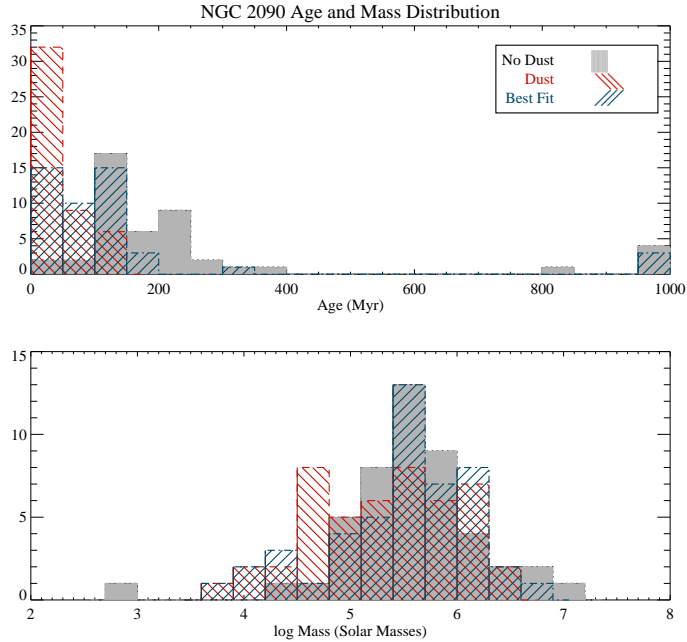}
\caption{The age and mass distribution for NGC 2090, with conventions as in Figure~\ref{fig:0628age}.}
\label{fig:2090age}
\end{figure}

\begin{figure}[H]
\epsscale{0.7}\plotone{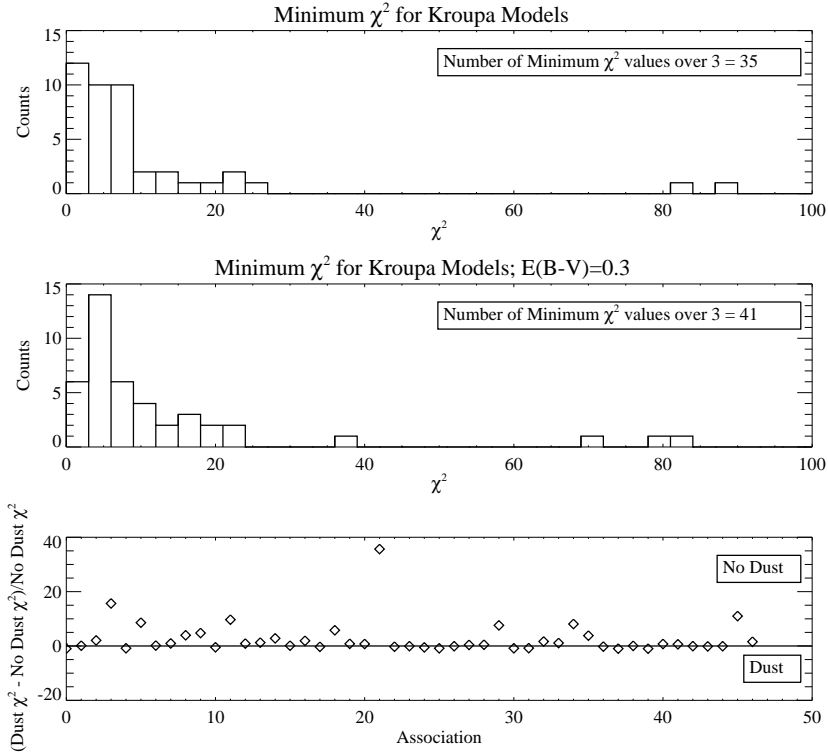}
\caption{NGC 2090 - As with NGC 0628, from a straight analysis of the minimum $\chi^2$ values, this galaxy appears to have low dust content except for a few associations.}
\label{fig:2090chi}
\end{figure}

The portion of NGC 2090 imaged contains 47 star forming associations.  Table~\ref{tbl:2090} shows the break-down of associations by type, the average and median age and mass of all associations, and the age dispersion.  Its dust map (not shown) shows about half of its stellar complexes are not associated with any dust, while the other half, hugging close to the spiral arms, may be.  However, the percentage of associations with inverse correlations or better $\chi^2$ fits with dust is similar to that of NGC 0628 and these may just be chance superpositions.  Figure~\ref{fig:2090age} shows the age and mass distributions and Figure~\ref{fig:2090chi} shows the $\chi^2$ distribution.  The no dust case has lower minimum $\chi^2$ by 64$\%$ on average.  12 associations ($\sim$26$\%$) show an inverse correlation between age and mass with no dust, with 5 ($\sim$11$\%$) showing both an inverse correlation and a better $\chi^2$ with dust.  13 ($\sim$28$\%$) associations have a better $\chi^2$ with dust. 8 ($\sim$13$\%$) associations are ionizing.

\begin{deluxetable}{ccccccc}
  \tabletypesize{\scriptsize}
  \tablecaption{NGC 2090 Association Statistics}
  \tablewidth{0pt}
  \tablehead{
  \colhead{ } &
  \colhead{Ionizing} &
  \colhead{Non-Ionizing} &
  \colhead{Evolved}  &
  \colhead{Average (Median)} &
  \colhead{log Average (Median)} &
  \colhead{Age}  \\
  \noalign{\smallskip}
  \colhead{ } &
  \colhead{($<$ 10 Myr) } &
  \colhead{(10 Myr$-$1 Gyr)} &
  \colhead{($>$ 1 Gyr) } &
  \colhead{Age [Myr]} &
  \colhead{Mass [$\Msun$]} &
  \colhead{Range [Myr]}
  }
  \startdata
  No Dust & 1 ($\sim$2$\%$)& 44 ($\sim$93$\%$) & 2 ($\sim$4$\%$) & 283 (191) &  5.57 (5.62)  & 2$-$1000 \\
  Dust & 23 ($\sim$48$\%$) & 24 ($\sim$51$\%$) & 0 & 38 (15) & 5.29 (5.39) & 1$-$216 \\
  Best Fit & 8 ($\sim$17$\%$) & 39 ($\sim$82$\%$) & 0 & 141 (96) & 5.48 (5.47) & 1$-$988
  \enddata
 \label{tbl:2090}
 \end{deluxetable}


\subsection{NGC 2841}

\begin{figure}[H]
\epsscale{0.7}\plotone{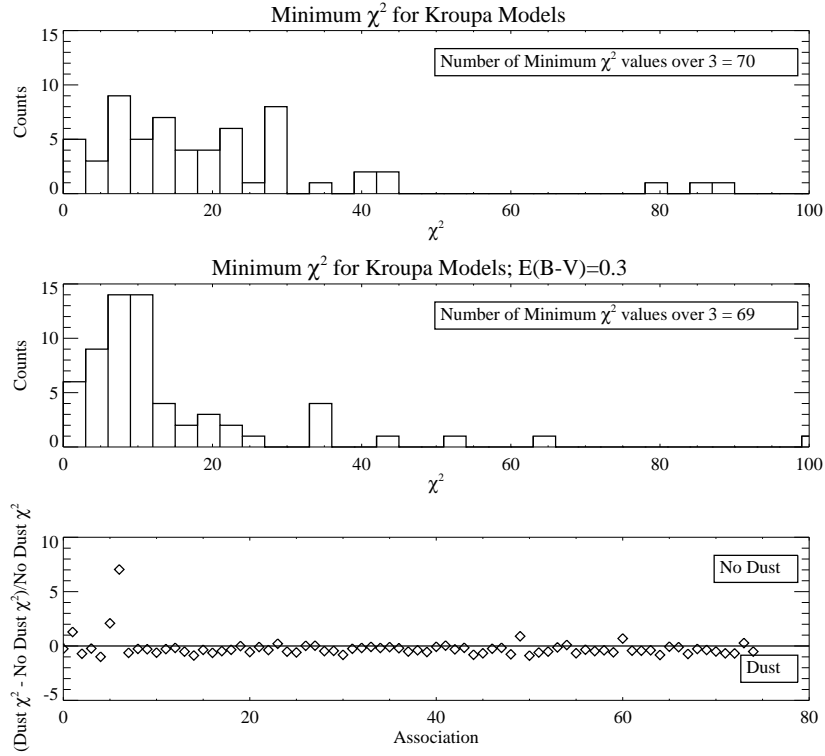}
\caption{NGC 2841 - Associations in NGC 2841 show better fits in the case of $E(B-V)$ = 0.3, both in the overall minimum $\chi^2$ values (top, middle) and in the strengths of the improvement for individual associations (bottom).  This is further supported by the number of inverse correlations between age and mass in the no dust case.}
\label{fig:2841chi}
\end{figure}

NGC 2841 has 75 associations in the portion imaged by IRAC and is by far the most likely candidate of our galaxy sample to have relatively large amounts of dust.  23 ($\sim$31$\%$) associations show an inverse correlation between age and mass with no dust.  19 of those also show a decrease in the minimum $\chi^2$ (Figure~\ref{fig:2841chi}) and an additional 45 ($\sim$60$\%$) associations show this same decrease with a positive correlation.  The associations that show some sign of dust also have excess 8.0 $\mu$m emission in all cases except five, where the 8.0 $\mu$m point is missing.  The dust map (Figure~\ref{fig:dust} (right)) shows quite a few complexes very obviously associated with dust emission and the minimum $\chi^2$ value improves with dust by 36$\%$ on average.

Due to the relative importance of dust in this galaxy, we questioned (as mentioned above) whether a different dust model may yield a significant change in our results.  So far we have assumed a simple model of Milky Way-type dust extinction which places all of the dust in front of the star forming region.  To test this, we also apply a dust attenuation model \citep{cal00}, which places the dust within the region, and find ages intermediate between the no dust and dust extinction case.  However, we also find that the overall fits are $\sim$30$\%$ worse in the dust attenuation case as measured by the minimum $\chi^2$ value.  

\begin{figure}[H]
\epsscale{0.6}\plotone{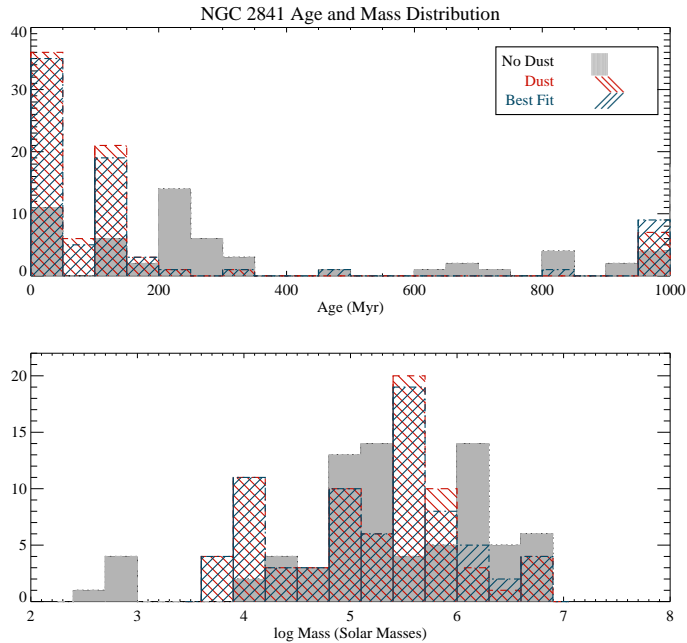}
\caption{As in Figure~\ref{fig:0628age}, the ages and mass distribution for NGC 2841.}
\label{fig:2841age}
\end{figure}

Table~\ref{tbl:2841} shows the break-down of the associations.  The average age with dust is $\sim$150 Myr, which is in good agreement with the average ages from the previous two galaxies and the median age of associations in M83 \citep{don08}.  This age doesn't change much when the best fits are taken into account.  The age range, however, still spans our models.  The age and mass distributions can be seen in Figure~\ref{fig:2841age}.  19 ($\sim$25$\%$) associations are ionizing.

\begin{deluxetable}{ccccccc}
  \tabletypesize{\scriptsize}
  \tablecaption{NGC 2841 Association Statistics}
  \tablewidth{0pt}
  \tablehead{
  \colhead{ } &
  \colhead{Ionizing} &
  \colhead{Non-Ionizing} &
  \colhead{Evolved}  &
  \colhead{Average (Median)} &
  \colhead{log Average (Median)} &
  \colhead{Age}  \\
  \noalign{\smallskip}
  \colhead{ } &
  \colhead{($<$ 10 Myr) } &
  \colhead{(10 Myr$-$1 Gyr)} &
  \colhead{($>$ 1 Gyr) } &
  \colhead{Age [Myr]} &
  \colhead{Mass [$\Msun$]} &
  \colhead{Range [Myr]}
  }
  \startdata
  No Dust & 5 ($\sim$6$\%$)& 52 ($\sim$70$\%$) & 18 ($\sim$24$\%$) & 496 (295) &  5.36 (5.25)  & 1$-$1000 \\
  Dust & 21 ($\sim$28$\%$) & 54 ($\sim$72$\%$) & 0 & 150 (56) & 5.18 (5.44) & 1$-$998 \\
  Best Fit & 19 ($\sim$25$\%$) & 56 ($\sim$74$\%$) & 0 & 190 (56) & 5.20 (5.44) & 1$-$988
  \enddata
 \label{tbl:2841}
 \end{deluxetable}

\newpage

\subsection{NGC 3621}

\begin{figure}[H]
\epsscale{0.6}\plotone{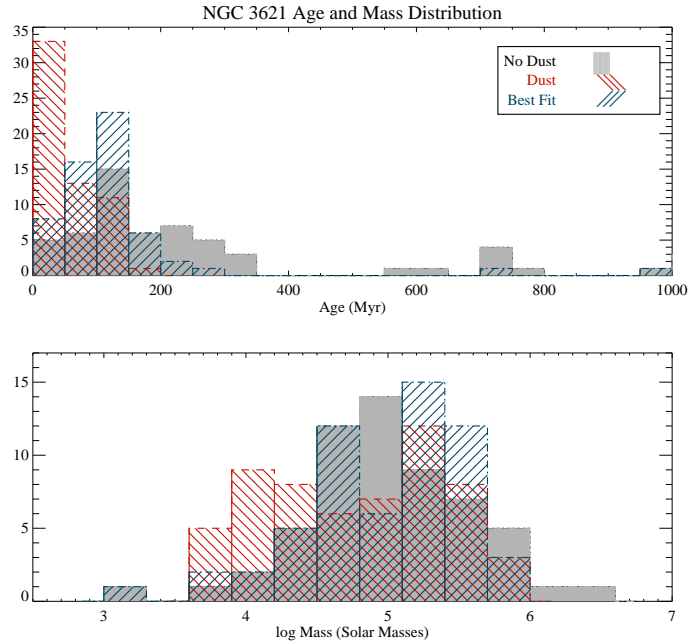}
\caption{As in Figure~\ref{fig:0628age}, the ages and mass distribution for NGC 3621. }
\label{fig:3621age}
\end{figure}

\begin{figure}[H]
\epsscale{0.7}\plotone{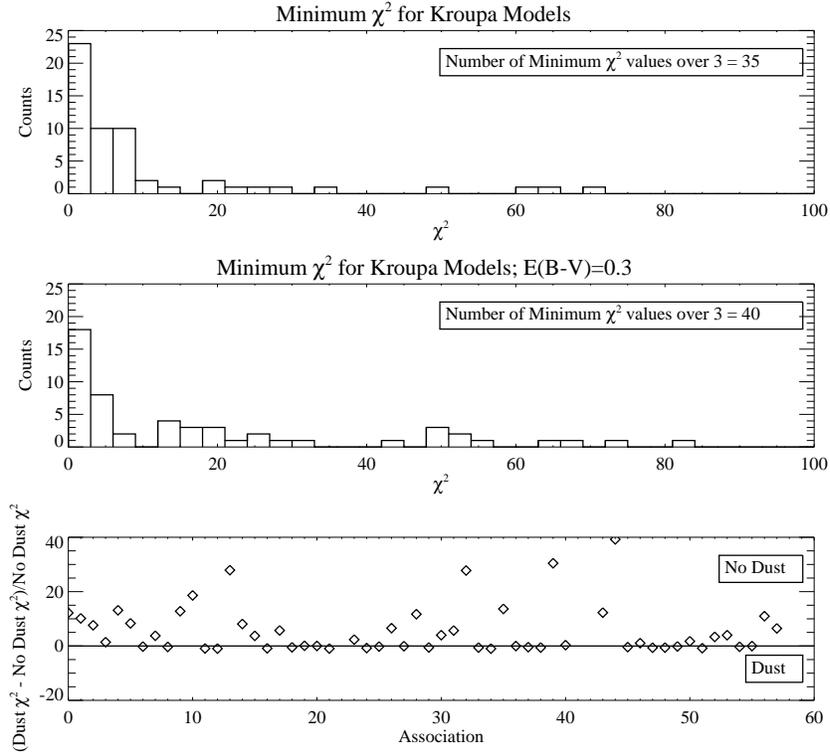}
\caption{NGC 3621 - Like NGC 0628 and NGC 2090, the $\chi^2$ distribution for NGC 3621 does not indicate a strong dust presence.}
\label{fig:3621chi}
\end{figure}

NGC 3621 contains 58 associations in the IRAC FOV and is our closest galaxy and deepest imaging.  It displays very little signs of dust, with only 5 ($\sim$9$\%$) associations with a inverse correlation between age and mass and 20 ($\sim$34$\%$) associations where dust improves the minimum $\chi^2$.  Table~\ref{tbl:3621} shows the association break-down and the $\chi^2$ improvement is 174$\%$ in the no dust case.  The age and mass distributions can be seen in Figure~\ref{fig:3621age} and $\chi^2$ distribution in Figure~\ref{fig:3621chi}.  4 ($\sim$6$\%$) associations are ionizing.

\begin{deluxetable}{ccccccc}
  \tabletypesize{\scriptsize}
  \tablecaption{NGC 3621 Association Statistics}
  \tablewidth{0pt}
  \tablehead{
  \colhead{ } &
  \colhead{Ionizing} &
  \colhead{Non-Ionizing} &
  \colhead{Evolved}  &
  \colhead{Average (Median)} &
  \colhead{log Average (Median)} &
  \colhead{Age}  \\
  \noalign{\smallskip}
  \colhead{ } &
  \colhead{($<$ 10 Myr) } &
  \colhead{(10 Myr$-$1 Gyr)} &
  \colhead{($>$ 1 Gyr) } &
  \colhead{Age [Myr]} &
  \colhead{Mass [$\Msun$]} &
  \colhead{Range [Myr]}
  }
  \startdata
  No Dust & 1 ($\sim$2$\%$)& 54 ($\sim$93$\%$) & 3 ($\sim$5$\%$) & 278 (191) &  4.98 (4.99)  & 6$-$1000 \\
  Dust & 27 ($\sim$46$\%$) & 31 ($\sim$53$\%$) & 0 & 47 (20) & 4.79 (4.93) &  1$-$160 \\
  Best Fit & 4 ($\sim$7$\%$) & 54 ($\sim$93$\%$) & 0 & 129 (100) & 4.96 (5.13) & 1$-$988
  \enddata
 \label{tbl:3621}
 \end{deluxetable}

\newpage

\subsection{NGC 5055}

\begin{figure}[H]
\epsscale{0.6}\plotone{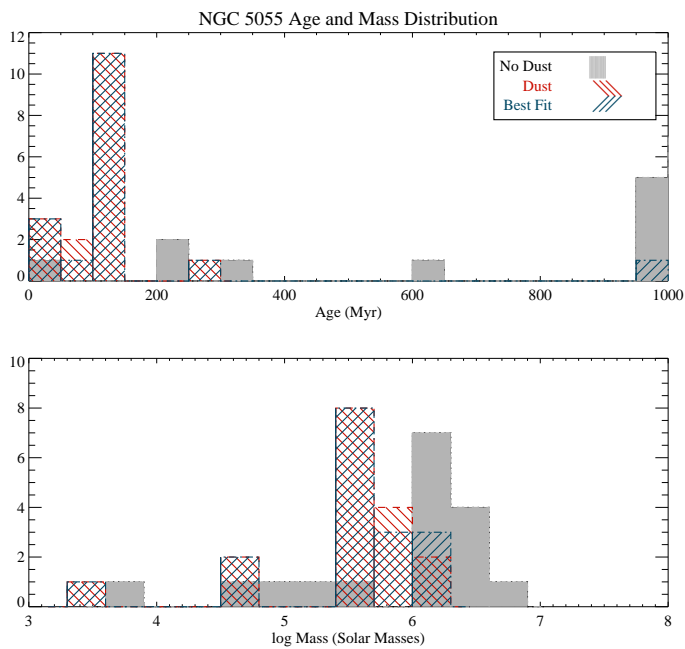}
\caption{As in Figure~\ref{fig:0628age}, the ages and mass distribution for NGC 5055.}
\label{fig:5055age}
\end{figure}

\begin{figure}[H]
\epsscale{0.7}\plotone{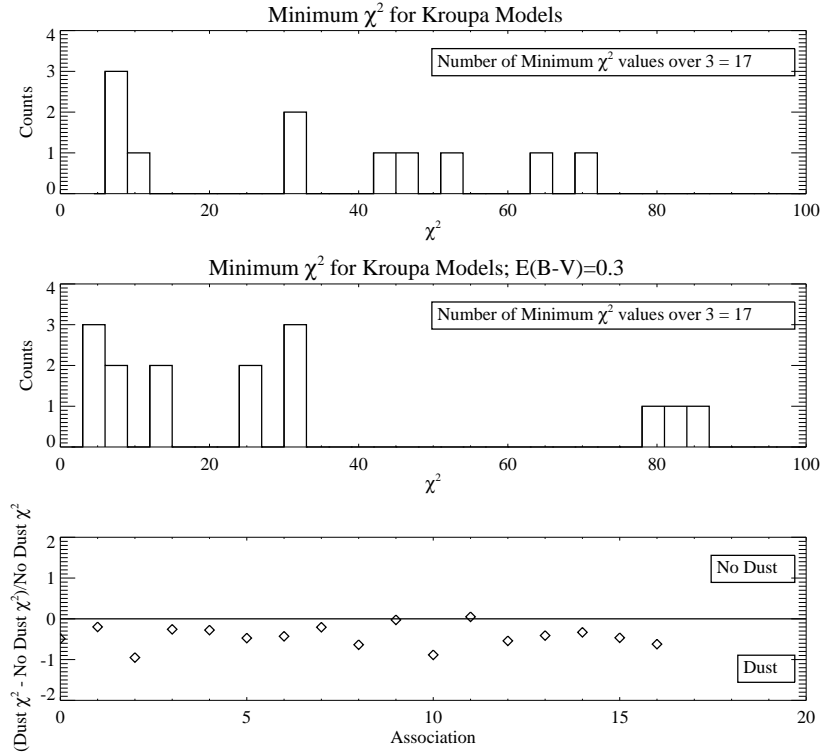}
\caption{NGC 5055 - While the same evidence that led us to suppose that the outer regions of NGC 2841 are relatively dusty compared to our other galaxies is present here, it is hard to tell due to the small number of associations and the large amount of noise in the NGC 5055 UV images.}
\label{fig:5055chi}
\end{figure}

While 146 associations were detected in the region imaged in NGC 5055, the FUV and NUV images are intrinsically faint and contain high levels of background noise spikes, which SExtractor had difficulty separating from actual sources.  As such, noise spikes had to be eliminated by hand before removing foreground stars and background galaxies.  The final conservative association count is 17, with fully half of those classified as evolved without dust.  As can be seen in Table~\ref{tbl:5055}, the average ages in all three categories is higher than for previous galaxies.  The age and mass distributions can be seen in Figure~\ref{fig:5055age} and the $\chi^2$ distribution in Figure~\ref{fig:5055chi}, with a $\chi^2$ improvement in the dust case of 43$\%$.  Only 1 ($\sim$6$\%$) association is ionizing.

\begin{deluxetable}{ccccccc}
  \tabletypesize{\scriptsize}
  \tablecaption{NGC 5055 Association Statistics}
  \tablewidth{0pt}
  \tablehead{
  \colhead{ } &
  \colhead{Ionizing} &
  \colhead{Non-Ionizing} &
  \colhead{Evolved}  &
  \colhead{Average (Median)} &
  \colhead{log Average (Median)} &
  \colhead{Age}  \\
  \noalign{\smallskip}
  \colhead{ } &
  \colhead{($<$ 10 Myr) } &
  \colhead{(10 Myr$-$1 Gyr)} &
  \colhead{($>$ 1 Gyr) } &
  \colhead{Age [Myr]} &
  \colhead{Mass [$\Msun$]} &
  \colhead{Range [Myr]}
  }
  \startdata
  No Dust & 0 & 10 ($\sim$58$\%$) & 7 ($\sim$41$\%$) & 786 (989) &  5.85 (6.11)  & 35$-$1000 \\
  Dust & 1 ($\sim$6$\%$) & 16 ($\sim$94$\%$) & 0 & 106 (116) & 5.45 (5.72) & 4$-$998 \\
  Best Fit & 1 ($\sim$6$\%$) & 16 ($\sim$94$\%$) & 0 & 159 (116) & 5.47 (5.57) & 4$-$988
  \enddata
 \label{tbl:5055}
\end{deluxetable}

\subsection{Cumulative Properties}  

The cumulative statistics of these five galaxies have demonstrated that star forming associations in the outer disks are represented by a large age dispersion with a median age of $\sim$100 Myr and an average age of $\sim$158 Myr (Table~\ref{tbl:all}).  Though extinction has a non-negligible effect on the age determinations of any individual association, the large percentage of young associations that are nevertheless beyond the ionizing stage within the extinction range assumed here is consistent with the presence of UV emission and the scarcity of H$\alpha$ emission, \citep{thi05,gil05,god10}.  These associations also show results consistent with low dust and low extinction in outer disk, although, again, more in-depth studies are needed to determine the dust content of individual associations.

\begin{figure}[H]
\epsscale{0.6}\plotone{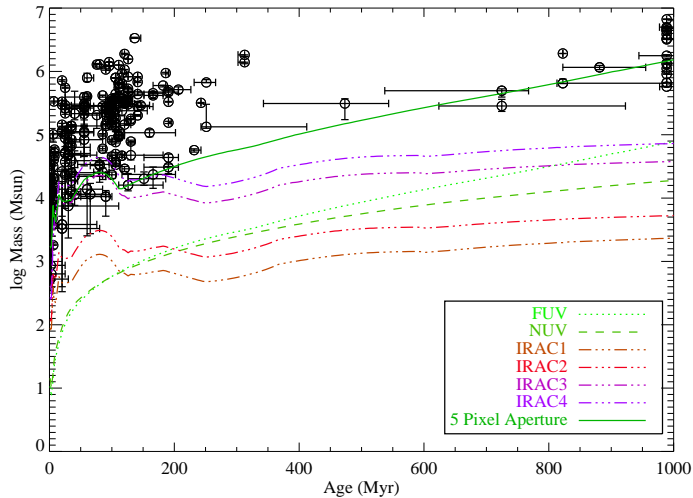}
\caption{Logarithmic mass ($\Msun$) as a function of the age (Myr) of all star forming associations.  The dotted and dashed lines indicate our per pixel detection limits, as determined by the 2$\sigma$ sensitivity in each band.  The limiting sensitivity for a 5 pixel aperture is also indicated.  For ages $\lesssim$50 Myr, the minimum detectable cluster mass is set by the 3.6$\mu$m sensitivity; for ages $\gtrsim$50 Myr, the limit is set by the FUV.}
\label{fig:sel}
\end{figure}

The selection effects due to the detections limits in all bands can be seen in Figure~\ref{fig:sel}.  The dominant effect is that we begin to lose sensitivity to associations below $\sim$10$^4$ $\Msun$ for ages older than $\sim$50 Myr due to the FUV sensitivity.  In addition, $\sim$25$\%$ of our associations fall into low signal to noise regions in the IRAC coverage maps, which affects our completeness as we may not have selected very young (blue) associations in these regions with high UV flux but undetected 3.6 $\mu$m emission.   

\begin{figure}[H]
\epsscale{0.6}\plotone{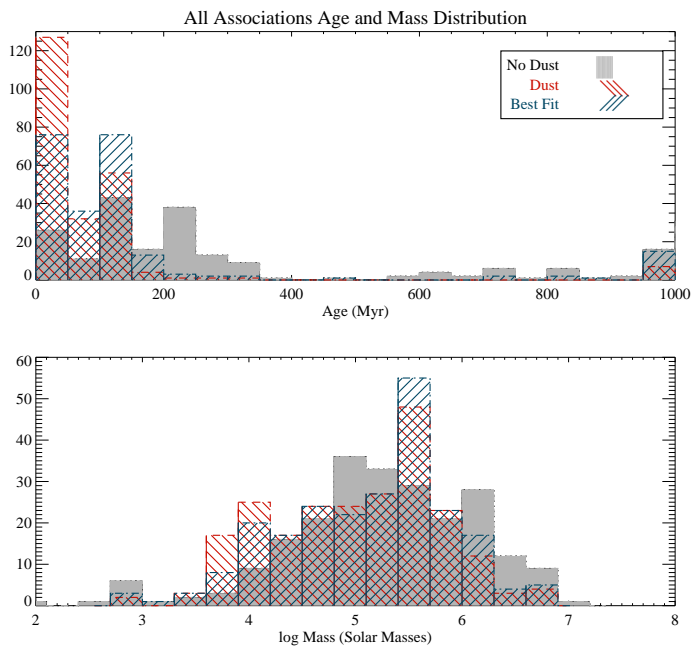}
\caption{This histogram includes all associations.  From this we can see that the median age is $\sim$100 Myr and the range of ages is populated. Two things should be noted here: 1) Associations beyond $\sim$200 Myr are underrepresented due to selection bias.  2) Associations categorized as 1 Gyr have undetermined ages due to the range of our models (1 Myr$-$1 Gyr).  In addition, due to the lack of models at older ages, these associations are represented in the `best fit' category as dusty associations as dust reddens the SEDs and provides a better fit to old (red) associations.}
\label{fig:allage}
\end{figure}

The age and masses of all associations in this study can be seen in Figure~\ref{fig:allage}, which demonstrates the wide range of ages seen in the star forming associations in outer disks.  Unfortunately, the incompleteness of our sample means that the two most prominent features in this histogram -- the peak at $\sim$100 Myr and the paucity of associations beyond $\sim$200 Myr -- are best explained as selection effects due to the fact that sources were originally selected to be FUV bright.  As the FUV mainly traces the youngest stages of SF, this selects heavily for ages $<$ 100 Myr and heavily against associations $>$ 200 Myr.  A simple Poissonian analysis of the 100 Myr peak shows it to be at the 6$\sigma$ level, indicating that in the age range to which we are most sensitive star forming associations are preferentially on the older end.  However, studies have shown that age determinations based on integrated colors can result in age pile-ups \citep{hun03, fal05}. A more complete sample is needed before any determination about the star formation history of outer disks can be made.   It is also important to note that since our models stop at 1 Gyr, associations that are listed as 1 Gyr actually have undetermined ages and, in the `best fit' category, they are preferentially selected as dusty, due to the fact that the presence of dust yields younger ages and we lack old-age (red) models without dust. Finally, none of our galaxies are undergoing a major merger, but there is evidence for interactions with small companions which may provide regular triggering of star formation. 

\begin{deluxetable}{ccc}
  \tabletypesize{\scriptsize}
  \tablecaption{Total Association Statistics}
  \tablewidth{0pt}
  \tablehead{
  \colhead{} &
  \colhead{Average (Median)} &
  \colhead{log Average (Median)} \\
  \noalign{\smallskip}
  \colhead{ } &
  \colhead{Age [Myr]} &
  \colhead{Mass [$\Msun$]} 
  }
  \startdata
  No Dust & 391 (216) & 5.88 (5.23) \\
  Dust & 82 (25) & 5.59 (5.14) \\
  Best Fit & 158 (100) & 5.57 (5.26)
  \enddata
 \label{tbl:all}
\end{deluxetable}   

\subsection{Comparison with H$\alpha$ Observations}

Though most of the galaxies in this study have been imaged in H$\alpha$ as part of the Spitzer Infrared Nearby Galaxies Survey (SINGS; \citet{ken03}), only NGC 3621 has observations extending to the outer disk regions.  We performed a simple spatial comparison to determine if our stellar clusters are associated with H$\alpha$ emission.  39 (66$\%$) of the associations have no H$\alpha$ detections while the other 19 (34$\%$) are spatially coincident with H$\alpha$ detections ranging from strong point sources to nearby diffuse emission.  Associations that have observed H$\alpha$ emission are generally younger with a median age for $E(B-V)$ = 0.0 ($E(B-V)$ = 0.3) of 100 (6) Myr for those with, and 231 (80) Myr for those without.  Due to the large uncertainties in our age determinations, this is consistent with our results. It is also important to note due to our selection criteria (using the FUV to select stellar clusters and then requiring IRAC data and other criteria to reduce contaminants), we do not claim a complete sample and thus cannot make a fair comparison with the number of H$\alpha$ sources.  A statistical sample of H$\alpha$ emitting sources across several XUV disks would allow us to extend this study and further evaluate the nature of outer disk star-forming clusters.


\section{IMF Analysis and Discussion}

In previous sections, we determined the age and masses of star forming association in outer disks for two extreme values of extinction with a fixed Kroupa IMF and found that at both extremes, there exists a wide age dispersion with a median age older than that at which ionizing stars will still exist.  We now address how a difference in the IMF will affect our age determinations.

There are two possible modifications to the IMF that could be explored that would be consistent with the observed lack of H$\alpha$: either the IMF can be truncated at some value, suggesting that there is some mechanism preventing high mass stars from forming \citep{pfl09,wei04,wei05}, or the slope of the IMF can be steepened.  Due to the coarse timesteps of our models and the assumed ages of our associations (beyond the ionizing stage), we believe that we are not sensitive to the differences between truncation and steepening and we choose here to test our ages against steeper IMF slopes.  The effect on our models should be that fewer massive stars are generated, resulting in less ionizing/UV emission and redder theoretical SEDs.  Then our associations should be fit at younger ages.

We fit our associations with two more Starburst99 model sets with $\alpha$ = 3.3 \citep{meu09} and $\alpha$ = 4.5 for the mass range 0.5$-$100 $\Msun$.  All other parameters are held constant (see Section 3).  The result is consistent with the chance in IMF slope having little effect on the age determinations made previous in this study.  On average, associations actually became older by 26 Myr for $\alpha$ = 3.3 and 36 Myr for $\alpha$ = 4.5, in the case of no dust.  The case of $E(B-V)$ = 0.3 also resulted in a small change to older ages.  As a sanity check, an IMF was generated with an extreme slope of 6.0 for a smaller range (1-400 Myr).  These fits did result in much younger ages, as predicted.  However, slopes this extreme have no observational evidence and are often disregarded in the literature.  

\begin{figure}[H]
\epsscale{1}\plottwo{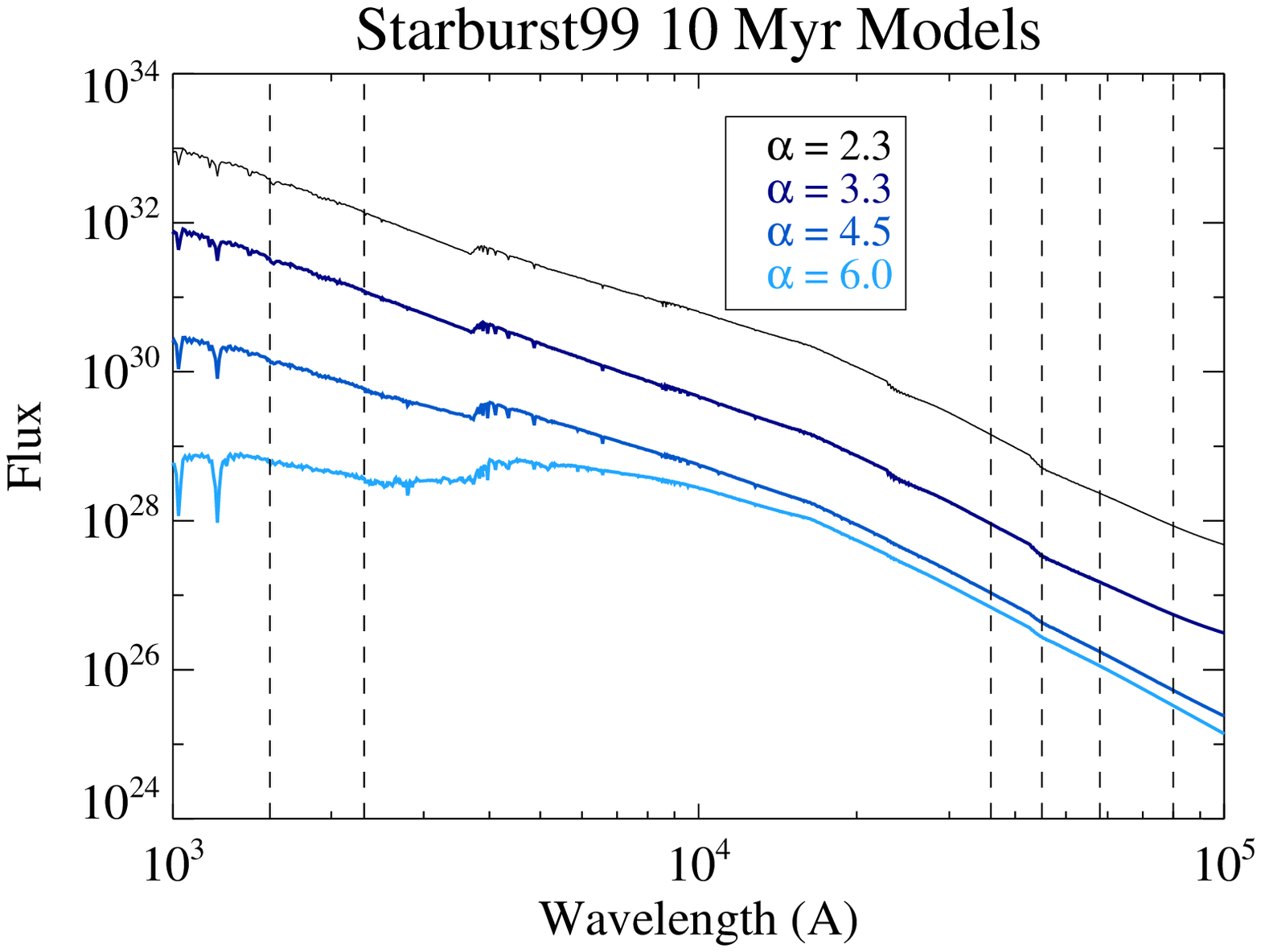}{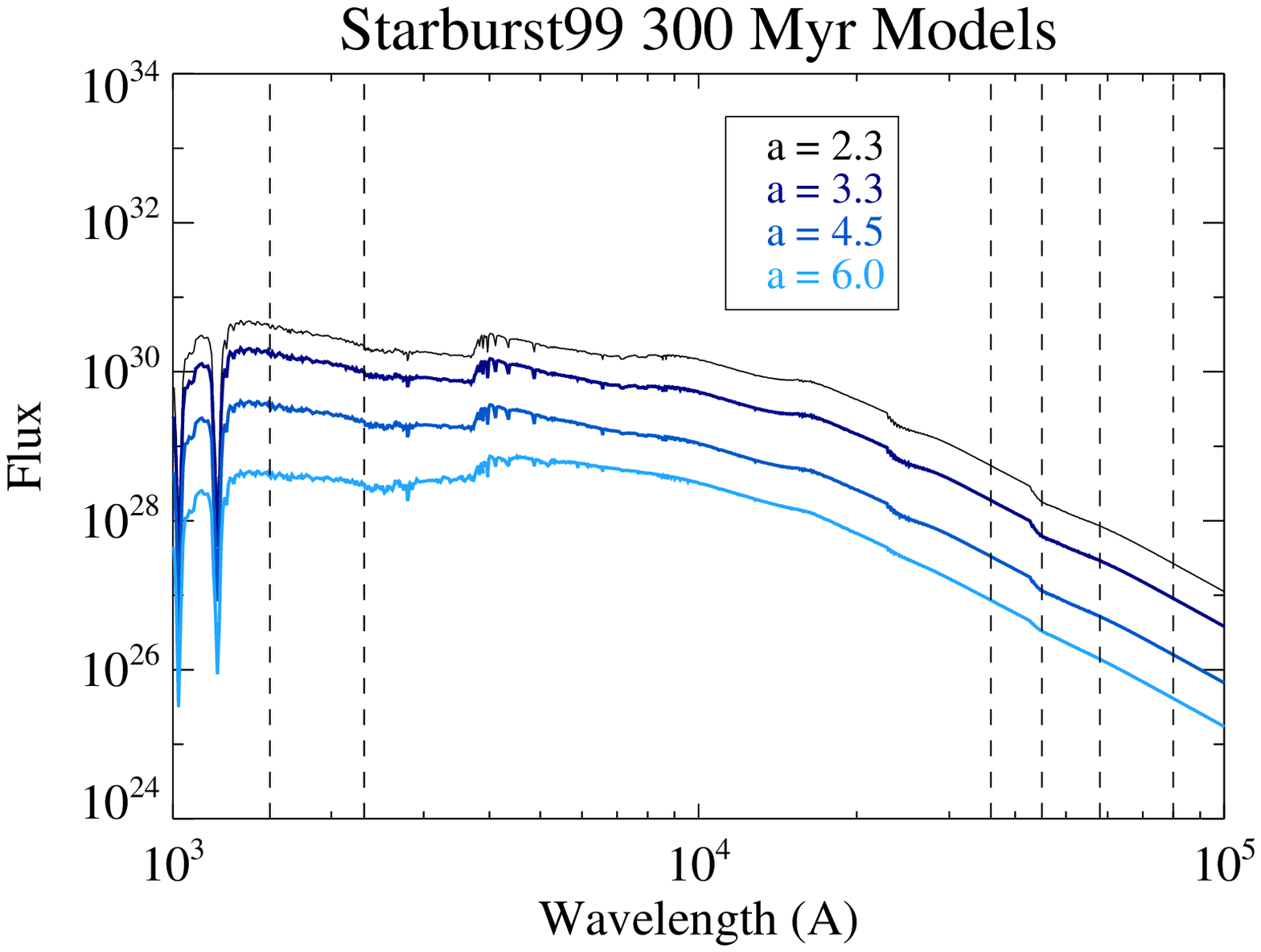}
\caption{Two Starburst99 age models, 10 Myr (left) and 300 Myr (right), are shown for four IMF slopes: $\alpha$ = 2.3, 3.3, 4.5, and 6.0.  The six dashed lines are the (from left to right) FUV, NUV, 3.6 $\mu$m, 4.5 $\mu$m, 5.8 $\mu$m, and 8.0 $\mu$m bands.  When using a long wavelength baseline, the age is determined by the relative strength between the UV and IR emission.  For the slopes of 2.3$-$4.5, the ratio between UV and IR emission is relatively constant, meaning that the age of a star forming association is largely independent of the IMF slope.  This is contrasted by an IMF with a slope of 6.0, which has a very different ratio.  The offset between slopes does affect the mass of the associations. }
\label{fig:star}
\end{figure}

\begin{figure}[H]
\epsscale{0.6}\plotone{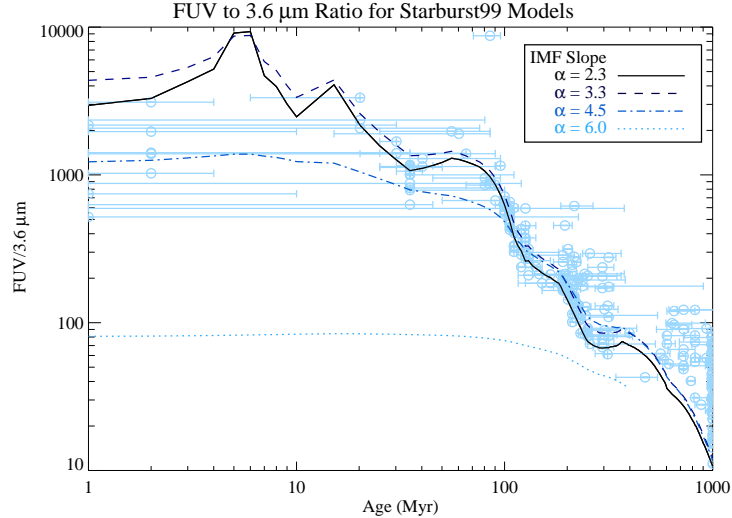}
\caption{The FUV$-$3.6 $\mu$m color as a function of age for the sets of models with IMF slope $\alpha$ = 2.3 (solid), 3.3 (dashed), 4.5 (dot-dash) and 6.0 (dotted).  The FUV$-$3.6 $\mu$m color for all associations in this study is shown in light blue for ages based on an assumed Kroupa ($\alpha$ = 2.3) IMF.  The typical FUV$-$3.6 $\mu$m error bar is smaller than the datapoints. }
\label{fig:FUV36}
\end{figure}

The relative insensitivity of our best fit ages to the steepening of the stellar IMF slope can be understood by recalling that a steeper IMF will have the most immediate effect on the short-lived massive stars, but not on the stars contributing to the emission in the GALEX and Spitzer IRAC bands.  This can be seen in Figure~\ref{fig:star}, where the spectral energy distribution of the 10~Myr and 300~Myr Starburst99 cluster models are plotted for a range of IMF slopes.  IMF slopes up to $\alpha$ = 4.5 produce minimal impact on the cluster's spectral energy distribution.  As an additional test, we show in Figure~\ref{fig:FUV36} the FUV$-$3.6$\mu$m color as a function of model age for each of the four IMF slopes overplotted with the FUV$-$3.6 $\mu$m color for all star forming associations using the best fit ages as determined with a Kroupa IMF.   The vast majority of associations occupy the space where the FUV$-$3.6 $\mu$m colors overlap for IMF slopes 2.3, 3.3, and 4.5.

\begin{figure}[H]
\epsscale{1}\plottwo{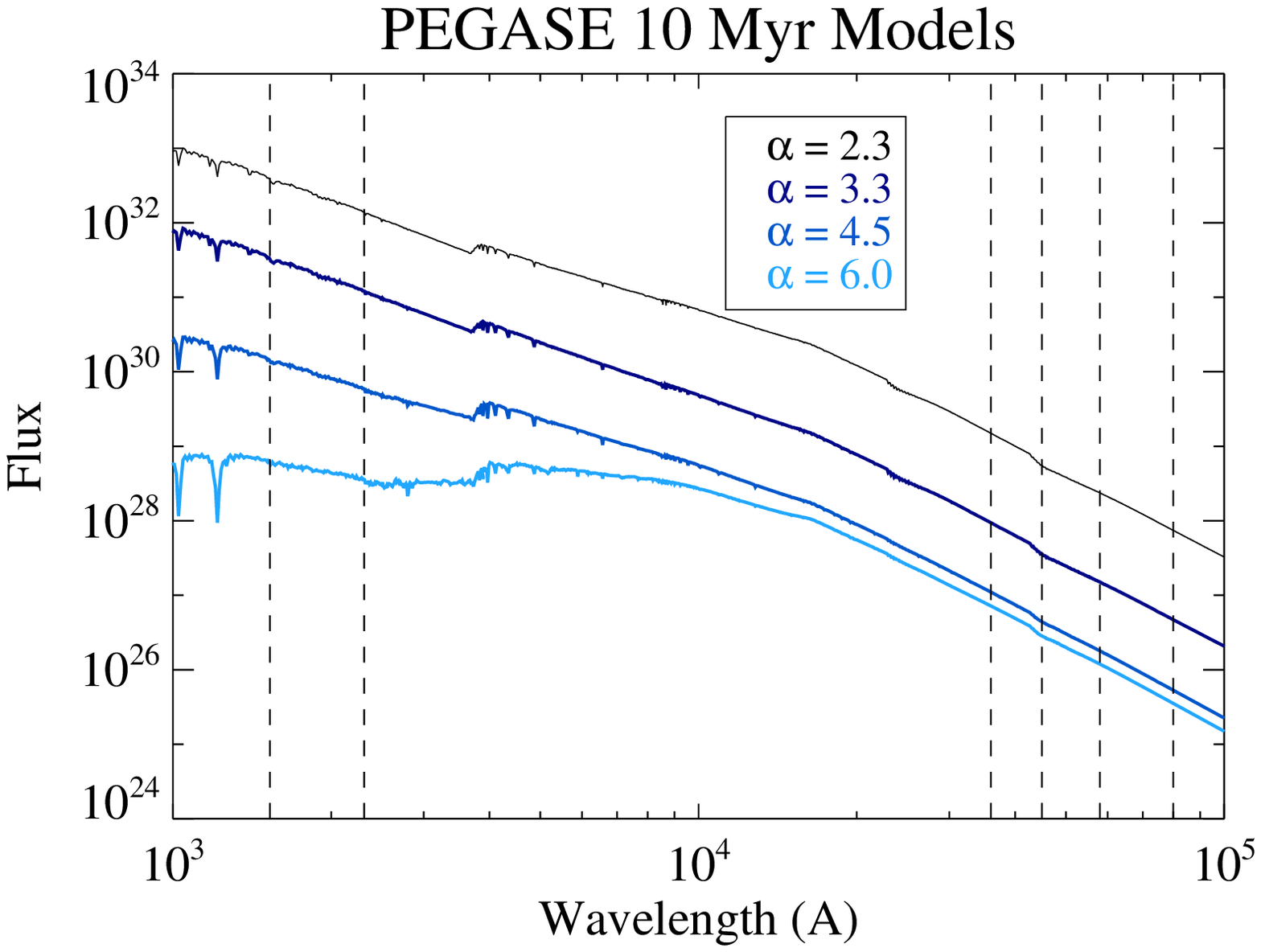}{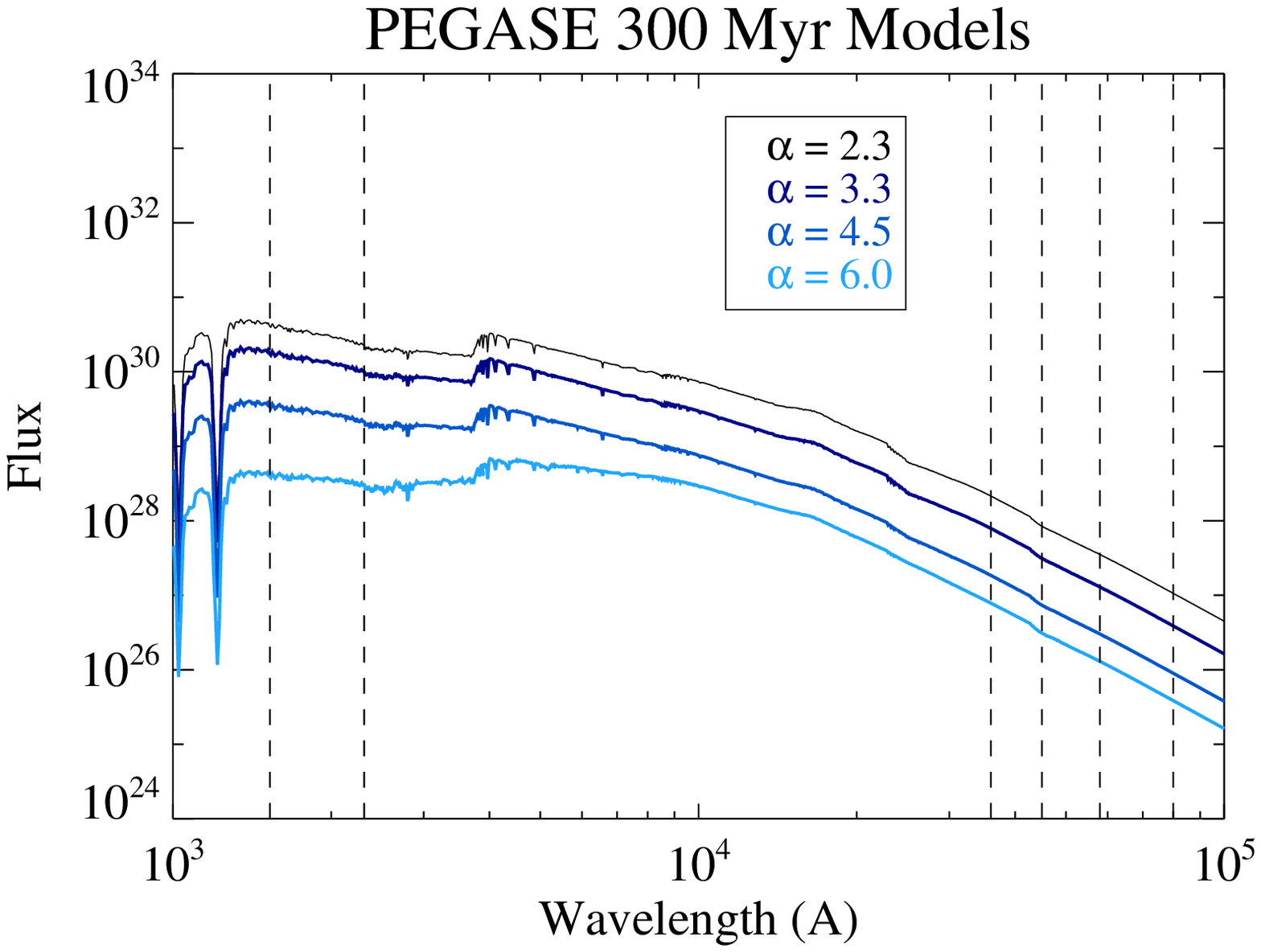}
\caption{As in Figure~\ref{fig:star}, with PEGASE II models.  The same effect is observed, even though the models do have subtle differences.}
\label{fig:peg}
\end{figure}

To further determine that this is not some systematic feature of these particular models, we also generated PEGASE II models \citep{fio99} for the same slopes in the same mass range at the same metallicity for 10 My and 300 Myr.  This is shown in Figure~\ref{fig:peg}.  For both sets of models, it is clear even by eye that the ratio between the UV and the IR bands remains relatively constant from slope 2.3 to 4.5 as compared to that same ratio for a slope of 6.0.  Since the age of an association is determined by the shape of the SED across these wavebands, this means that the age predicted is largely independent of the IMF slope.  This in effect breaks the age-IMF degeneracy for our long wavelength baseline SEDs and tells us that, assuming we correctly account for extinction and have accurate models (the two models, Starburst99 and PEGASE II, shown here will give different age results due to differences in the predicted MIR bands), we are looking at the real age distribution.  Of course, just because we cannot distinguish between IMF slopes does not mean that a top-light IMF does not exist in these regions. Future studies to quantify the dearth of H$\alpha$ relative to UV and an accurate age distribution can be used to constrain IMF properties.  At this point, however, while a non-standard IMF cannot be excluded, it is not required to account for the observed properties of our associations.  The dearth of H$\alpha$ emission in the outer disk regions is well accounted for by the spread in ages of star forming associations.  That some associations are as old as, or exceed, 1 Gyr of age, suggests that the original triggering mechanism of the in-situ star formation has caused sufficient instability in the outer disk that gas clouds are still effectively collapsing and forming stars to the present time.

\section{Conclusions and Future Work}

In this paper, we have leveraged Spitzer IRAC imaging against the GALEX FUV and NUV wavebands to determine the ages and masses of star forming associations in the low-density XUV disks of five nearby spiral galaxies.  These associations likely consist of a few stellar clusters with typical cumulative masses of $\sim$10$^{5.3}$ $\Msun$.  The age dispersion across all five disks is large, spanning the range 1 Myr$-$$\gtrsim$1 Gyr, with a median age of $\sim$100 Myr.  Though the age determinations can vary considerably for different extinction values, the wide age dispersion and the old median age are robust against both extinction (in the range $E(B-V)$ = 0$-$0.3 mag) and steepening of the IMF slope.  This indicates that the age dispersion is real and sufficient to explain the observed dearth of H$\alpha$ in XUV disks as compared to UV emission without the need to invoke a non-standard IMF.  

Our insensitivity to the IMF slope is potentially useful in constraining the IMF properties in these regions.  The next step in this study is to obtain deep H$\alpha$ imaging of these regions, analogous to \citet{god10}, which can be used to confirm our results and give us further insight into the youngest modes of star formation in these low-density regions.  In addition, optical datapoints would allow us to further constrain the ages of these associations.  In particular, the U$-$B color is highly sensitive to the age of an association while being relatively insensitive to extinction.  Combined, H$\alpha$ observations and accurate ages would provide a powerful diagnostic tool for studies of the local IMF.    


\acknowledgments{This work has made use of the Spitzer Space Telescope data obtained as part of the Cycle 3 GO program 30753. It has been partially supported by the JPL, Caltech, Contract Number 1296193. This research has made use of the NASA/IPAC Extragalactic Database (NED) which is operated by the Jet Propulsion Laboratory, California Institute of Technology, under contract with the National Aeronautics and Space Administration.}





\end{document}